\documentclass[ALICE,manyauthors]{cernphprep}
\usepackage{hyperref}
\usepackage{graphics}
\usepackage{multirow}
\usepackage{perpage}
\usepackage{endnotes}
\MakePerPage{footnote}
\newcommand {\pT}{\ensuremath{p_{\mathrm{T}}} }
\newcommand {\meanpT}{\ensuremath{\langle p_{\mathrm{T}}} \rangle} 

\newcommand {\mT}{\ensuremath{m_{\mathrm{T}}}}

\newcommand {\rap} {\mbox{$\left | y \right | $}}
\newcommand {\pseudorap} {\mbox{$\left | \eta \right | $}}

\newcommand {\dNdy}{\mathrm{d}N/\mathrm{d}y }

\newcommand {\dEdx}{\mathrm{d}E/\mathrm{d}x }
\newcommand {\PbPb}{\mbox{Pb--Pb} }
\newcommand{\ppbar} {\mbox{$\mathrm {p\overline{p}}$}}

\newcommand {\m} {\, \mbox{${\rm m}$}}

\newcommand {\stat}{({\it stat.})~}
\newcommand {\syst}{({\it syst.})~}
\newcommand {\mass} {\mbox{\rm MeV$\kern-0.15em /\kern-0.12em c^2$}}
\newcommand {\tev} {\mbox{${\rm TeV}$}}
\newcommand {\gev} {\mbox{${\rm GeV}$}}
\newcommand {\mev} {\mbox{${\rm MeV}$}}

\newcommand {\mom} {\mbox{\rm GeV$\kern-0.15em /\kern-0.12em c$}}
\newcommand {\gmom} {\mbox{\rm GeV$\kern-0.15em /\kern-0.12em c$}}
\newcommand {\mmass} {\mbox{\rm MeV$\kern-0.15em /\kern-0.12em c^2$}}
\newcommand {\gmass} {\mbox{\rm GeV$\kern-0.15em /\kern-0.12em c^2$}}
\newcommand {\mmom} {\mbox{\rm MeV$\kern-0.15em /\kern-0.12em c$}}

\newcommand {\musec} {\mbox{$\mu {\rm s}$}}
\newcommand {\nsec} {\mbox{${\rm ns}$}}
\newcommand {\psec} {\mbox{${\rm ps}$}}
\newcommand {\fm} {\mbox{${\rm fm}$}}
\newcommand {\cm} {\mbox{${\rm cm}$}}
\newcommand {\mm} {\mbox{${\rm mm}$}}
\newcommand {\mim} {\mbox{$ \mu {\rm m}$}}

\newcommand{\pbar} {\mbox{$\mathrm {\overline{p}}$}}

\newcommand{\Kzs}{\mbox{$\mathrm {K^0_S}$}}
\newcommand{\rmLambda}{\mbox{$\mathrm {\Lambda}$}}
\newcommand{\rmAlambda}{\mbox{$\mathrm {\overline{\Lambda}}$}}
\newcommand{\rmXi}{\mbox{$\mathrm {\Xi^{-}}$}}
\newcommand{\rmAxi}{\mbox{$\mathrm {\overline{\Xi}^{+}}$}}
\newcommand{\Xis}{\mbox{$\mathrm {\Xi^{-}+\overline{\Xi}^{+}}$}}

\newcommand{\Jpsi} {\mbox{J\kern-0.05em /\kern-0.05em$\psi$}}

\begin{document}
\begin{titlepage}
\PHnumber{2010-065}                 
\PHdate{6 December 2010}              

\title{Strange particle production in proton-proton collisions
 	at $\sqrt{s}=0.9$ TeV with ALICE at the LHC}
\ShortTitle{Strange particle production in pp at $\sqrt{s}=0.9$ TeV}
%
%
\Collaboration{ALICE Collaboration%
         \thanks{See Appendix~\ref{app:collab} for the list of collaboration 
                      members}\\[1cm]
                 This publication is dedicated to the memory of our colleague Ren\'{e} Kamermans}
\ShortAuthor{ALICE Collaboration}      
\begin{abstract}
The production of mesons containing strange quarks ($\Kzs$, $\phi$) and both singly and doubly
strange baryons ($\rmLambda$, $\rmAlambda$,  and $\Xis$) are measured at central rapidity
in pp collisions at $\sqrt{s}$~=~0.9~$\tev$ with the ALICE experiment at the LHC.
The results are obtained from the analysis of about 250~k minimum bias events recorded in 2009. 
Measurements of yields (dN/dy) and transverse momentum spectra at central rapidities for inelastic
pp collisions are presented.
For mesons, we report yields ($\langle \dNdy \rangle$) of $0.184 \pm 0.002~\stat \pm 0.006~\syst$
for $\Kzs$ and  $0.021 \pm 0.004~\stat \pm 0.003~\syst$ for $\phi$.
For baryons, we find $\langle \dNdy \rangle = 0.048 \pm 0.001~\stat  \pm 0.004~\syst$ for
$\rmLambda$, $0.047 \pm 0.002~\stat \pm 0.005~\syst$ for $\rmAlambda$ and
$0.0101 \pm 0.0020~\stat \pm 0.0009~\syst$ for $\Xis$.
The results are also compared with predictions for identified particle spectra from QCD-inspired
models and provide a baseline for comparisons with both future pp measurements at higher energies
and heavy-ion collisions.
\end{abstract}
\end{titlepage}
\setcounter{page}{2}
%
%
\section{Introduction}
\label{intro}

The production of hadrons at high transverse momenta in high energy proton-proton
collisions is reasonably well described by perturbative Quantum Chromodynamics  (pQCD)
in terms of hard parton-parton scattering (large momentum transfers) followed by 
fragmentation~\cite{Sjostrand:2006za,Engel:1995sb}.
However, the low-momentum region, where most particles are produced and which
therefore contributes most to the underlying event, is dominated by soft interactions.
In the soft regime, it has been found that particle production can be described effectively
by models based on emission from an equilibrated system at a specific temperature
and baryo-chemical potential, with additional accounting of conserved 
quantities~\cite{Becattini:1997rv,Kraus:2008fh,Becattini:2009ee}.
It can also be treated in the framework of QCD inspired phenomenological models,
that include multi-parton processes, extrapolated to very low-momentum transfers~\cite{Sjostrand:2004ef}.
The contribution and evolution of multi-parton processes as a function of $\sqrt{s}$
is difficult to establish.
Measurements of identified particles at the beam injection energy of the LHC and in the low 
transverse momentum ($\pT$) region, along with their comparison with QCD-inspired
models, constitute a baseline for comparisons with higher centre-of-mass energies.
The low $\pT$ cutoff achievable through the low material budget, low central barrel 
magnetic field (0.5 T) and excellent particle identification (PID) of the ALICE detectors, allows
an accurate measurement of the low momentum region at mid-rapidity.

The differential transverse momentum yields ($\pT$ spectra) and integrated yields at
central rapidity of $\Kzs$, $\phi$, $\rmLambda$, $\rmAlambda$ and $\Xis$
have been measured by the ALICE experiment during the commissioning phase of the LHC
(December 2009)~\cite{Evans:2008zzb} with the very first proton-proton
collisions~\cite{Collaboration:2009dt} and are reported in this article.
A sample of 250~k minimum bias pp collisions at $\sqrt{s}$~=~0.9~$\tev$ has been selected
with triggers combining several fast detectors~\cite{Aamodt:2010ft}. 
Measurements are performed using the tracking devices and the main PID
detectors of ALICE in the central rapidity region ($\rap<0.8$).
A comparison of the transverse momentum shapes (mass dependence and
mean transverse momentum) with PYTHIA~\cite{Sjostrand:2006za} and PHOJET~\cite{Engel:1995sb}
is provided.

This article is organized as follows.
Section 2 presents the experimental conditions, the minimum bias event selection as well
as a brief description of the main detectors and the associated event reconstruction tools
used for the analysis.
Section 3 is dedicated to the data analysis, including track and topological selections, signal
extraction methods and the corresponding efficiency corrections.
The determination of the systematic uncertainties are also described in this section.
In section 4, the $\pT$ spectra and the integrated yields of the studied particle species are
given and compared with previous measurements and model predictions.
Conclusions are given in section 5. 
%
%
%
%
\section{Experimental set-up and data collection}
\label{sec:setup}
A detailed description of the ALICE experimental setup and its detector subsystems
can be found in~\cite{Aamodt:2008zz}.
\subsection{Main detectors and reconstruction techniques used for the analyses}
\label{sec:detrec}
The central barrel of ALICE covers polar angles from $45^{\circ}$  to $135^{\circ}$ over the
full azimuth.
It is embedded in the large L3 solenoidal magnet providing a nominal magnetic field $B$
of $0.5~\mathrm{T}$. 
Within the barrel, the two tracking detectors used in these present analyses consist of an
Inner Tracking System (ITS), composed of 6 cylindrical layers of high-resolution Silicon
detectors  and a cylindrical Time Projection Chamber (TPC).
PID is performed using secondary (displaced) vertex reconstruction, invariant mass
analysis and single track PID methods, which include the measurement of specific
ionization in the ITS and the TPC, and the information from the Time-Of-Flight detector (TOF).  
\subsubsection{The Inner Tracking System}
\label{sec:detits}
The Silicon Pixel Detector (SPD) corresponds to the two innermost ITS layers.
These two layers have a very high granularity with a total of about $9.8$ million pixels,
each with a size of $50 \times 425~\mim^2$.
They are located at radii of $3.9$ and $7.6~\cm$ and the pseudorapidity coverages are
$|\eta|<2.0$ and $|\eta|<1.4$ respectively.
The detector provides a position resolution of $12~\mim$ in the r$\phi$ direction and
about $100~\mim$ in the direction along the beam axis.
It can also deliver a signal for the first level of trigger (L0) in less than $850~\nsec$.
The two layers of the Silicon Drift Detector (SDD), located at radii of $15.0$ and $23.9~\cm$,
are composed of 260 sensors, including 133 000 collection anodes
with a pitch of $294~\mim$.
They provide a charge deposit measurement and a position measurement with a resolution
of about $35~\mim$ in the r$\phi$ direction and about $25~\mim$ in the beam
direction~\cite{:2010ys}.
The Silicon Strip Detector (SSD) consists of $1698$ double-sided sensors (with a strip
pitch of $95~\mim$ and a stereo angle of $35$ mrad) arranged in 2 layers located at radii
of $38$ and $43~\cm$.
It provides a measurement of the charge deposited in each of the $2.6$ million strips, as well
as a position measurement with a resolution of $20$~$\mim$ in the r$\phi$ direction and
about $800~\mim$ in the beam direction. 

The ITS sensor modules were aligned using survey information and
tracks from cosmic-ray muons and pp collisions. The corresponding methods 
are described in~\cite{:2010ys}.

The percentage of operational channels in the ITS during the 2009 run is $82\%$
for the SPD, $91\%$ for the SDD and $90\%$ for the SSD.
\subsubsection {The Time Projection Chamber}
\label{sec:dettpc}
The ALICE TPC is a cylindrical drift detector with a pseudorapidity coverage of $|\eta|\leq 0.9$~\cite{Alme:2010}.
It has a field cage filled with $90~\rm{m}^3$ of Ne/CO$_2$/N$_2$ (85.7/9.5/4.8\%). 
The inner and outer radii of the active volume are of $85~\cm$ and $247~\cm$ respectively
and the length along the beam direction is $500~\cm$.   
Inside the field cage, ionization electrons produced when charged particles traverse
the active volume on either side of the central electrode (a high voltage membrane
at $-100~\mathrm{kV}$) migrate to the end plates in less than $94~\musec$.
A total of 72 multi-wire proportional chambers, with cathode pad readout, instrument
the two end plates of the TPC which are segmented in 18 sectors and amount to a
total of $557,568$ readout pads.
The ALICE TPC ReadOut (ALTRO) chip, employing a $10$ bit ADC at $10~\mathrm{MHz}$
sampling rate and digital filtering circuits, allows for precise position and linear energy loss
measurements with a gas gain of the order of $10^4$.

The position resolution in the $r\phi$ direction varies from $1100~\mim$ to $800~\mim$ 
when going from the inner to the outer radius whereas the resolution along the beam
axis ranges between $1250~\mim$ and $1100~\mim$.  
\subsubsection {The Time-Of-Flight detector}
\label{sec:dettof}
The ALICE Time-Of-Flight detector~\cite{Akindinov:2010zz} is a cylindrical assembly
of Multi-gap Resistive Plate Chambers (MRPC) with an inner radius of $370~\cm$ and an
outer radius of $399~\cm$, a pseudorapidity range $|\eta| < 0.9$ and full azimuth angle, 
except for the region $260<\phi<320$ at $\eta$ near zero where no TOF modules were
installed to reduce the material in front of the Photon Spectrometer.
The basic unit of the TOF system is a 10-gap double-stack MRPC strip $122~\cm$ long
and $13~\cm$ wide, with an active area of $120\times7.4~\cm^2$ subdivided into two
rows of 48 pads of $3.5\times2.5~\cm^2$. 
Five modules of three different types are needed to cover the full cylinder along the z direction.
All modules have the same structure and width ($128~\cm$) but differ in length. 
The overall TOF barrel length is $741~\cm$ (active region).
It has 152,928 readout channels and an average thickness of $25-30\%$ of a radiation
length, depending on the detector zone.
For pp collisions, such a segmentation leads to an occupancy smaller than $0.02~\%$.
Its front-end electronics is designed to comply with the basic characteristics of a MRPC
detector, i.e. very fast differential signals from the anode and cathode readout.
Test beam results demonstrated a time resolution below $50~\psec$, dominated by the jitter
in the electronic readout.
\subsubsection {The VZERO Counters}
\label{sec:vzero}
The VZERO counters are two scintillator hodoscopes located along the beam direction
at $-0.9~\m$ and $3.3~\m$ from the geometrical centre of the experiment.
They correspond to a coverage of $-3.7<\eta<-1.7$ and $2.8<\eta<5.1$ respectively and
have a time resolution close to $0.5~\nsec$.
They are used as trigger detectors and help to remove beam-gas interaction background.
\subsubsection {Track reconstruction and particle identification}
\label{sec:recpid}
The global tracking system in the ALICE central barrel (combining the ITS and the TPC)
covers the pseudorapidity window $|\eta|<0.9$.

The reconstruction in the tracking detectors begins with charge cluster finding.  
The two coordinates of the crossing points (space points) between tracks  and detector
sensitive elements (pad rows in the TPC, and silicon sensors in the ITS) are calculated
as the centres of gravity of the clusters.
The errors on the space point positions are parametrized as functions of the cluster size
and of the deposited charge.
In the TPC, these errors are further corrected during the tracking, using the crossing angles
between tracks and the pad rows.

The space points reconstructed at the two innermost ITS layers (pixel
detector, SPD) are then used
for the reconstruction of the primary vertex.
One space point from the first SPD layer and one from the second layer are combined
into pairs called ``tracklets".
The primary vertex is consequently reconstructed in 3D as the location that minimizes
the sum of the squared distances to all the tracklet extrapolations.
If this fails, the algorithm instead reconstructs the $z$ coordinate of the vertex by correlating
the $z$ coordinates of the SPD space points, while for $x$ and $y$ the
average position of the beam in the transverse plane (measured basis by a
dedicated calibration procedure on a run-by-run basis) is assumed.

Track reconstruction in ALICE is based on the Kalman filter approach and is discussed
in detail in~\cite{Alessandro:2006yt}. 
The initial approximations for the track parameters (the ``seeds'') are constructed using
pairs of space points taken at two outer TPC pad rows separated by a few pad rows and
the primary vertex. 
The primary vertex position errors for this procedure are considered to be as big as
$3~\cm$.
The seeds for the secondary tracks are created without using the primary vertex,
since such a constraint would unnecessarily reduce the V0 finding efficiency.
The space points are searched along the line connecting the pairs of points taken
at those two outer TPC pad rows.

Once the track seeds are created, they are sorted according to the estimate
of their transverse momentum ($\pT$).
Then they are extended from one pad row to another in the TPC and from one layer
to another in the ITS towards the primary vertex. 
Every time a space point is found within a prolongation path defined by the current
estimate of the covariance matrix, the track parameters and the covariance matrix are
updated using the Kalman filter.
For each tracking step, the estimates of the track parameters and the covariance matrix
are also corrected for the mean energy loss and Coulomb multiple scattering in the 
traversed material.
The decision on the particle mass to be used for these corrections is based on the
$\dEdx$ information given by the TPC when available. 
If the information is missing or not conclusive, a pion mass is assumed.
Only five particle hypotheses are considered: electrons, muons, pions, kaons and protons.

All the tracks are then propagated outwards, through the ITS and the TPC.
When possible, they are matched with the hits reconstructed in the TOF detector. 
During this tracking phase, the track length and five time-of-flight hypotheses per track
(corresponding to the electron, muon, pion, kaon and proton masses) are calculated.
This information is later used for the TOF PID procedure.
The track parameters are then re-estimated at the distance of closest approach (DCA)
to the primary vertex applying the Kalman filter to the space points already attached.
Finally, the primary vertex is fitted once again, now using reconstructed tracks and the
information about the average position and spread of the beam-beam interaction region
estimated for this run.

In pp collisions, the track reconstruction efficiency in the acceptance of TPC 
saturates at about 90\% because of the effect of the dead zones between its sectors.
It goes down to about $75\%$ around  $\pT=1~\gmom$ and drops to $45\%$ at  $0.15~\gmom$.
It is limited by particle decays (for kaons), track bending at low $\pT$ and
absorption in the detector material.
The amount of material traversed by particles near $\eta=0$ is about 11\% of of a radiation length
including the beam pipe, the ITS and the TPC (with service and support).

The overall $\pT$ resolution is at least as good as the TPC-standalone resolution,
which is typically $1\%$ for momenta of $1~\gmom$ and $7\%$ for momenta of
$10~\gmom$, and follows the parameterization $(\sigma(\pT)/\pT)^2 = (0.01)^2 + (0.007~\pT)^2$
where $\pT$ is expressed in $\gmom$ (see~\cite{Aamodt:2010my} for the details).

The resolution of the track transverse impact parameter (the minimal distance between
a track and the primary vertex in the transverse plane) depends on the precision of track
and primary vertex reconstruction.
These in turn depend on the momentum, and, in the case of the vertex, on the number of
contributing tracks.
As it was estimated from the data, the transverse impact parameter resolution for a typical
pp event could be parameterized as 
$\sigma(\pT) = 50 + 60/(\pT)^{0.9}$ ($\sigma$ is in $\mim$, and $\pT$ is in $\gmom$), 
which was defined by the level of the ITS alignment achieved in 2009. 

The $\dEdx$ resolution of the TPC is estimated to be about $5\%$ for tracks
with 159 clusters~\cite{Alme:2010}, which is better than the design
value~\cite{Alessandro:2006yt}. When averaged over all reconstructed
tracks, this resolution is about 6.5\%.

During the run, the preliminary calibration of the TOF detector corresponds to
a resolution of $180~\psec$, which includes $140~\psec$ due to the jitter in
the absolute time of the collisions.
This contribution is reduced to about $85~\psec$ for those events with at least
3 tracks reaching the TOF, in which case an independent time zero determination 
is possible.
The matching efficiency with TPC tracks (which includes geometry, decays and
interaction with material) is on average $60\%$ for protons and pions and reaches
$65\%$ above $\pT = 1~\gmom$.
For kaons it remains sligthly lower~\cite{Alice:Pid}.
Above $\pT = 0.5~\gmom$, the TOF PID has an efficiency larger than 
$60\%$ with a very small contamination.
\subsection{LHC running conditions and triggers} 
\label{sec:lhcrun}
For the first collisions provided by the Large Hadron Collider, four low intensity
proton bunches ($10^9$ protons per bunch, giving the luminosity of the order
of $10^{26}$~cm$^{-2}$s$^{-1}$) per beam were circulated, and two pairs of
them crossed at the ALICE interaction point.
Under such conditions, the rate for multiple events in a given bunch-crossing
(``pile-up'') was negligible.
The energy in the centre of mass corresponded to twice the beam injection energy,
that is $\sqrt{s}$~=~$0.9~\tev$.
The data acquisition of ALICE was triggered by requiring two coincidence conditions:
i) the LHC bunch-crossing signal together with the two beam pick-up monitors (BPTX);
ii) ALICE minimum bias (MB) trigger requiring a combination of signals from the SPD
and from the VZERO counters.
For these analyses,  the MB$_{\rm{OR}}$ was used, which
is fulfilled when at least one of the VZEROs or the SPD trigger is fired~\cite{Aamodt:2010ft}.
The corresponding data rate was $\sim 10~\rm{Hz}$.
\section{Data analysis}
\subsection{Event and track selection} 
\label{event_and_track_selection}
\begin{figure}
\resizebox{0.5\textwidth}{!}{%
   \includegraphics{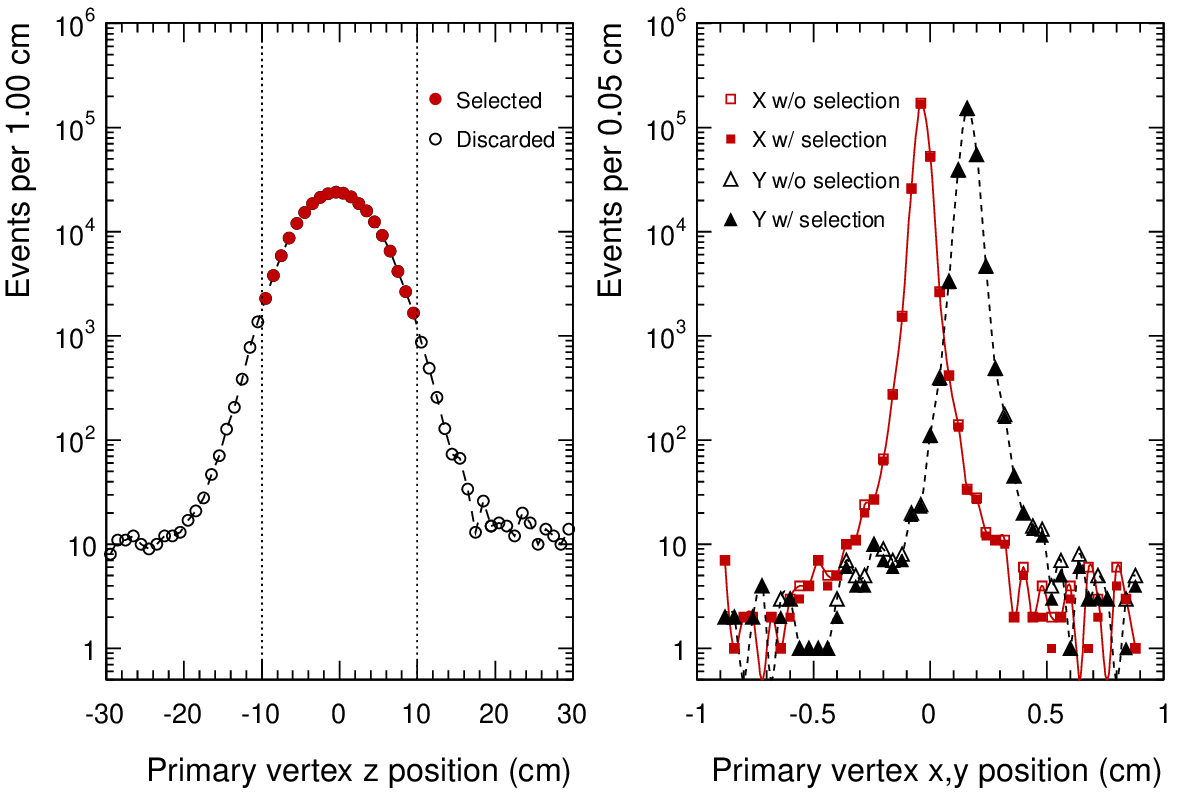}
}
\caption{Primary vertex distributions for the analysed events.
The left panel shows the distributions along the beam axis. Selected events (full symbols)
are required to have a reconstructed primary vertex with $|z|<10~\cm$.
The right panel corresponds to the directions perpendicular to the beam axis: horizontally
(i.e. $x$-direction, squares and full line) and vertically (i.e. $y$-direction, triangles and
dashed line).}
\label{fig:primvtx}
\end{figure}

The primary vertex is reconstructed using either SPD tracklets~\cite{Collaboration:2009dt} 
($5\%$ of the events) or global tracks ($95\%$ of the events).
Events are selected by requiring that the distance between the position of primary
vertex and the geometrical centre of the apparatus along the beam axis be less 
than $10~\cm$ ($\overline{z}=-0.40~\cm$ and $\mathrm{rms}_{z}= 4.24~\cm$,
where $\overline{z}$ is the average position of the primary vertex along the beam axis).
Events with less centred primary vertices ($|z|>10~\cm$) are discarded in order to minimize
acceptance and efficiency biases for tracks at the edge of the TPC detection
volume.
The average position and dispersion for both horizontal and vertical directions
are found to be $\overline{x}=-0.35~\mm$ ($\overline{y}=+1.63~\mm$)
and  $\mathrm{rms}_{x}= 0.23~\mm$ ($\mathrm{rms}_{y}=0.27~\mm$).
No conditions were applied on the $x$ and $y$ position of the vertex.
The total number of events used for obtaining the particle spectra and yields is
about $250$~k events.
Figure~\ref{fig:primvtx} shows the primary vertex distribution along the beam axis
(left panel) and for the $x$ and $y$ directions (right panel).
The dashed lines indicate the limits of the selected vertex region.

The normalization to the number of inelastic events (INEL) is obtained
in the same way as other ALICE analyses~\cite{Collaboration:2009dt,Alice:Pid}.
It leads to a correction for the normalization of $\sim 5\%$ with an uncertainty
of $2\%$.
This uncertainty is added to the ones described in section~\ref{subsec_sys} and mainly
related to the modeling of the fraction of diffractive events with several Monte Carlo
event generators.

Several quality criteria are defined for track selection.
Each track is required to have been reconstructed in the TPC in the initial outward-in
step of tracking and then successfully refitted in the final back-propagation to the primary
vertex as described in section~\ref{sec:recpid}.
It is also required that each track has at least 80 TPC clusters out of a maximum of
$159$.
At the reconstruction level, split tracks are rejected as well as those which may correspond
to daughters of kaons decaying in the TPC.
As the $\phi$ particle is a strongly decaying resonance, its daughters are indistinguishable
from primary particles at the reconstruction level and therefore primary track selections are used.
As a first step, each track is propagated to the reconstructed primary vertex.
If this operation is successful, the track is kept if it has a DCA smaller than $5~\mm$ ($3~\cm$)
in the transverse (longitudinal) direction with the additional constraints of having at least one SPD
cluster and a $\chi^2$ of less than 4 per cluster assignment (for each cluster, the $\chi^2$ has
two degrees of freedom).
%
%
\begin{table}
\caption{Track selection criteria.}
\label{tab:1}       
%
\begin{center}
\begin{tabular}{lcl}
\hline\noalign{\smallskip}
\hspace{0.5cm} Common selections &  &  \\
\noalign{\smallskip}\hline\noalign{\smallskip}
Detectors required for track rec./fit & & ITS,TPC \\ 
Number of TPC clusters $^{\rm a}$ & $>$ & $80$    \\
N($\sigma$) $\dEdx$ (TPC PID)     &  & $3~to~5$     \\
\noalign{\smallskip}\hline\noalign{\smallskip}
\hspace{0.5cm} Primary track selections &  &  \\
\noalign{\smallskip}\hline\noalign{\smallskip}
$\chi^{2}$ per cluster                        & $<$ & $4$     \\
DCA to primary vertex (r,z)       & $<$ & $(0.5,3.0)~\mathrm{cm}$   \\
Number of SPD clusters $^{\rm b}$ & $\geq$ & $1$ \\
\noalign{\smallskip}\hline\noalign{\smallskip}
\hspace{0.5cm} Secondary track selections &  &  \\
\noalign{\smallskip}\hline\noalign{\smallskip}
transverse momentum $^{\rm c}$  & $>$ & $160~\mmom$ \\
\noalign{\smallskip}\hline\noalign{\smallskip}
$^{\rm a}$ maximum number for the TPC is $159$; &  & \\
$^{\rm b}$ maximum number for the SPD is $2$; &  & \\
$^{\rm c}$ in the cases of $\Kzs$, $\rmLambda$ and $\rmAlambda$. &  & \\
\noalign{\smallskip}\hline
\end{tabular}
\end{center}
\end{table}

Depending on its lifetime, a particle may cross several layers of the ITS before weakly decaying.
The probability that the daughter tracks of $\Kzs$, $\rmLambda$, $\rmAlambda$ and $\Xis$ have
a hit in this detector decreases accordingly. 
Therefore, no specific condition on the number of ITS hits is required for the daughter tracks
of the reconstructed secondary vertices.
However, other quality criteria are applied for selecting the daughter tracks of weakly decaying
particles which are not considered as primaries. The selections described are summarized
in Table~\ref{tab:1}.
%
%
\begin{table}
\caption{Secondary vertex selection criteria.}
\label{tab:2}       
%
\begin{center}
\begin{tabular}{lcl}
\hline\noalign{\smallskip}
\hspace{0.5cm} Common selections &   &  \\
\noalign{\smallskip}\hline\noalign{\smallskip}
Minimum transverse decay radius &$>$& $0.2~\cm$   \\
Maximum transverse decay radius &$<$& $100~\cm$   \\
\noalign{\smallskip}\hline\noalign{\smallskip}
\hspace{0.5cm} $V0$ vertex selections ($\Kzs$, $\rmLambda$ and $\rmAlambda$) &  &  \\
\noalign{\smallskip}\hline\noalign{\smallskip}
DCA of V0 daugther track \\ to primary vertex &$>$& $0.05~\cm$   \\
DCA between V0 daughter tracks &$<$& $0.50~\cm$   \\
Cosine of V0 pointing angle ($\rmLambda$ and $\rmAlambda$)  &$>$& $0.99$  \\
\noalign{\smallskip}\hline\noalign{\smallskip}
\hspace{0.5cm} Cascade vertex selections &   &  \\ 
\noalign{\smallskip}\hline\noalign{\smallskip}
DCA of cascade daughter track \\ to primary vertex $^{\rm a}$       &$>$& $0.01~\cm$ \\
DCA between V0 daughter tracks    &$<$& $3.0~\cm$   \\
Cosine of V0 pointing angle       &$>$& $0.97$       \\
DCA of V0 to primary vertex          &$>$& $0.001~\cm$ \\
V0 invariant mass  &$>$& $1110~\mmass$ \\
V0 invariant mass  &$<$& $1122~\mmass$ \\
DCA between V0 and bachelor track &$<$& $3.0~\cm$   \\
Cosine of cascade pointing angle  &$>$& $0.85$ \\
\noalign{\smallskip}\hline\noalign{\smallskip}
$^{\rm a}$ for bachelor and each V0 daughter. &  & \\
\noalign{\smallskip}\hline
\end{tabular}
\end{center}
\end{table}
The measurement of differential yields in rapidity and $\pT$ bins cannot be performed
simultaneously for the particles considered due to the small available statistics.
Therefore the rapidity ranges are chosen such that i) the efficiency does not vary strongly for 
each species and ii) the rapidity distribution is sufficiently flat for it to be possible to rely on the
Monte Carlo to obtain the corrections.
\subsection{Particle reconstruction and identification methods} 
\label{particle_reco_and_pid}
\subsubsection{Topological reconstruction of $\Kzs$, $\rmLambda$, $\rmAlambda$ and $\Xis$}
The $\Kzs$, $\rmLambda$, $\rmAlambda$ and $\Xis$  are identified by applying selections
on the characteristics of their daughter tracks (see Table~\ref{tab:2}) and using their weak decay
topologies in the channels listed in Table~\ref{tab:3}.
%
%
\begin{table*}
\caption{Main characteristics of the reconstructed particles: valence quark content, mass, c$\tau$ and charged decay branching ratio (B.R.)~\cite{Nakamura:2010zzi}.}
\label{tab:3}       
\begin{center}
\resizebox{1.00\textwidth}{!}{
\begin{tabular}{llcclc}
\hline\noalign{\smallskip}
\multicolumn{2}{c}{Particles} &   mass ($\mmass$)   &  c$\tau$ & charged decay & B.R. (\%) \\
\noalign{\smallskip}\hline\noalign{\smallskip}
\multirow{2}{*}{Mesons}  & $\Kzs$ & 497.61  & $2.68~\cm$ & $\Kzs \rightarrow \pi^+ + \pi^-$ & 69.2 \\
              & $\phi$ ($s\bar{s}$) & 1019.46  & $45~\fm$ & $\phi \rightarrow \rm{K}^+ + \rm{K}^-$ & 49.2 \\
\noalign{\smallskip}\hline\noalign{\smallskip}
\multirow{2}{*}{Baryons}  & $\rmLambda$ ($uds$) and $\rmAlambda$ ($\overline{uds}$) & $1115.68$   & $7.89~\cm$ &  $\rmLambda \rightarrow p + \pi^-$ and $\rmAlambda \rightarrow \overline{p} + \pi^+$ & 63.9 \\
              & $\rmXi$  ($dss$) and  $\rmAxi$~($\overline{dss}$) & $1321.71$ & $4.91~\cm$ & $\rmXi \rightarrow  \rmLambda+ \pi^-$ and $\rmAxi \rightarrow \rmAlambda + \pi^+$ & 99.9   \\
\noalign{\smallskip}\hline
\end{tabular}
}
\end{center}
\end{table*}
%
%
%

The measurement of $\Kzs$, $\rmLambda$ and  $\rmAlambda$ is based on the 
reconstruction of the secondary vertex (V0) associated to their weak decay.  
The V0 finding procedure starts with the selection of secondary tracks, i.e.
tracks having a sufficiently large impact parameter with respect to the primary vertex. 
All possible combinations between two secondary tracks of opposite charge 
are then examined.
They are accepted as V0 candidates only if the DCA between them
is smaller than $0.5~\cm$.
The minimization of the distance between the tracks is performed numerically
using helix parametrizations in 3D. 
The V0 vertex position is a point on the line connecting the points of closest approach
between the two tracks. Its distance from each daughter track is taken
to be proportional to the precision of the track parameter estimations.
Once their position is determined, only the V0 candidates located inside a 
given fiducial volume are kept.
The inner boundary of this fiducial volume is at a radius of $0.2~\cm$ from the
primary vertex, while the outer limit is set at $100~\cm$.
Finally, for $\rmLambda$ and $\rmAlambda$ reconstruction, the V0 finding
procedure checks whether the particle momentum ($\vec{p}$) associated with the V0
candidate (calculated as the sum of the track momenta extrapolated to the position
of the DCA) points back to the primary vertex.  
This is achieved by applying a cut on the cosine of the angle (pointing angle 
$\theta_{\vec{p}}$) between $\vec{p}$ and a vector connecting the primary vertex
and the V0  position ($\cos{\theta_{\vec{p}}} > 0.99$).
The invariant mass of each candidate can then be calculated either under the
$\Kzs$ or the $\rmLambda$ hypothesis.

The TPC PID helps substantially to remove the combinatorial background for
the $\rmLambda$ and $\rmAlambda$ (mainly for the baryon daughter identification, 
while it is not needed for the $\Kzs$ decaying into pions).
TPC PID is described in paragraph~\ref{sec_phireco}.
The selections here concern the proton daughter only and have been chosen
to be looser for the daughter track with momentum below $0.7~\gmom$ ($\pm 5\sigma$) 
and tighter for higher momentum ($\pm 3\sigma$). 

The $\Xis$ particles are identified via their ``cascade'' decay topology.
The cascade finding procedure starts from the V0 finding procedure for the $\rmLambda$
daughter but with less stringent selection criteria (see Table~\ref{tab:2} and Cascade vertex
selections).
This is done to increase the efficiency and to allow for the fact  that the daughter
$\rmLambda$'s do not have to point back to the primary vertex.

The V0 candidates found within the $\rmLambda$ mass window ($1116 \pm 6~\mmass$)
are combined with all possible secondary tracks (bachelor candidates) with the exception
of both V0 daughter tracks.
A cut on the impact parameter of the bachelor track is applied to reject the primary particles 
which increase the combinatorial background.

A V0-bachelor association is performed if the distance of closest approach between the
bachelor track and the V0 trajectory (DCA between V0 and bachelor track) is small (less than
$3~\cm$).
Finally, this cascade candidate is selected if its reconstructed momentum points back to 
the primary vertex (cosine of cascade pointing angle).
The cascade finding is limited to the fiducial region used for V0 
reconstruction (see Table~\ref{tab:2}).

\begin{figure}
\resizebox{0.5\textwidth}{!}{%
  \includegraphics{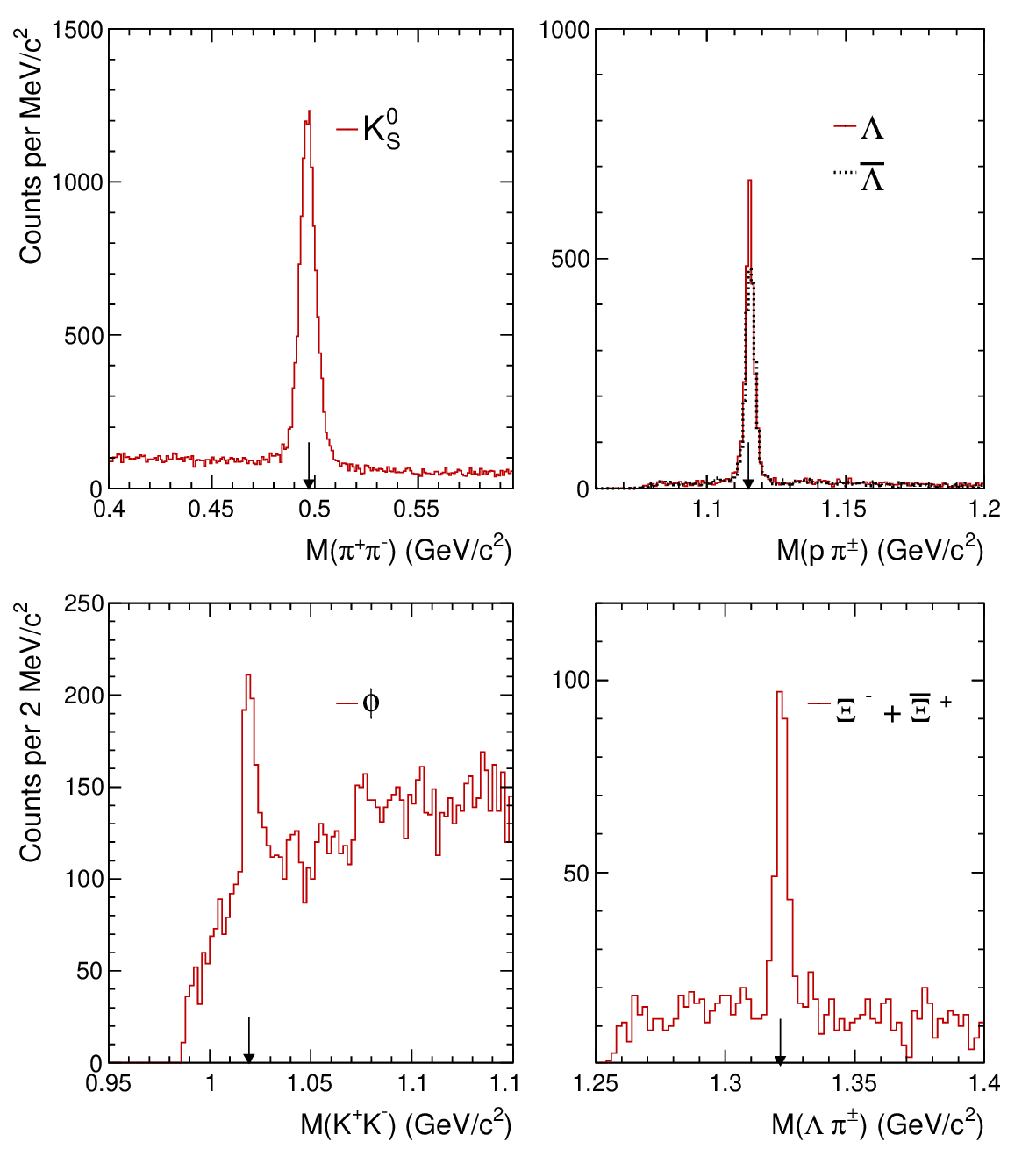}
}
\caption{Invariant mass distributions of $\Kzs$, $\rmLambda$ and $\rmAlambda$, $\phi$
and the sum $\Xis$. The vertical arrows indicate the nominal mass values from PDG.}
\label{fig:invmass}
\end{figure}

In addition to topological selections, the reconstruction of cascades also makes use of the
single-track PID information delivered by the TPC.
This is considered for each of the three daughters (both pions and the proton). 
For each track, a loose selection is required ($\pm 4\sigma$ over the whole momentum range) 
to reject the combinatorial background in part.
The resulting invariant mass distributions are presented in Fig.~\ref{fig:invmass}.

\subsubsection{Additional quality checks for $\Kzs$, $\rmLambda$, $\rmAlambda$}
\label{sec:qualchks}
\begin{figure}
\resizebox{0.5\textwidth}{!}{%
  \includegraphics{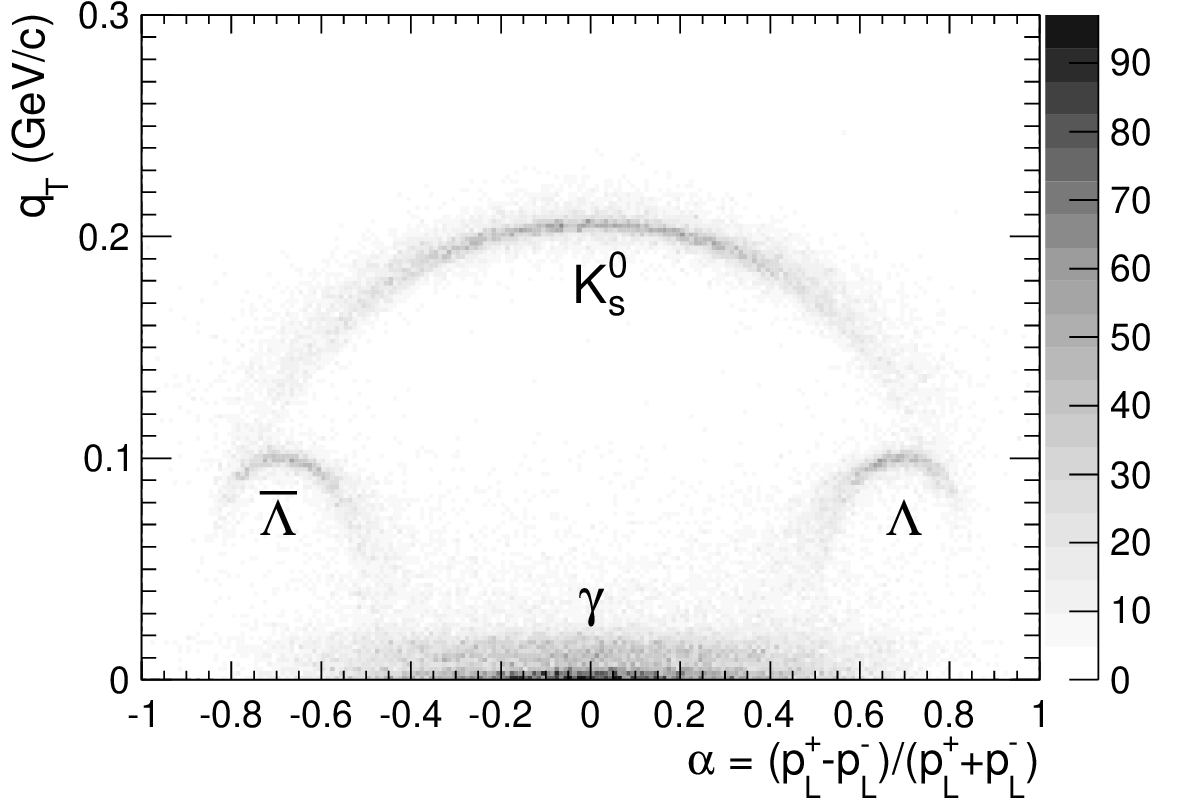}
}
\caption{Armenteros-Podolanski distribution for V0 candidates showing
a clear separation between $\Kzs$, $\rmLambda$ and $\rmAlambda$.
The $\gamma$ converting to $e^{+}e^{-}$ with the detector material are
located in the low $q_{T}$ region, where $q_{T}$ is the momentum component 
perpendicular to the parent momentum vector.}
\label{fig:armpodo}
\end{figure}

A significant fraction of the reconstructed V0 come from $\gamma$ conversion
in the detector material.
This can be clearly seen in the Armenteros-Podolanski distribution \cite{Armenteros} shown in 
Fig.~\ref{fig:armpodo} where $p_{\rm{L}}^{+}$ and $p_{\rm{L}}^{-}$ are the 
longitudinal components of the total momentum for the positive and negative
daughters respectively, relative to the direction of the V0 momentum vector.
The $\Kzs$, $\rmLambda$ and $\rmAlambda$ signal regions are symmetric and clearly distinguishable.

\begin{figure}
\resizebox{0.5\textwidth}{!}{%
  \includegraphics{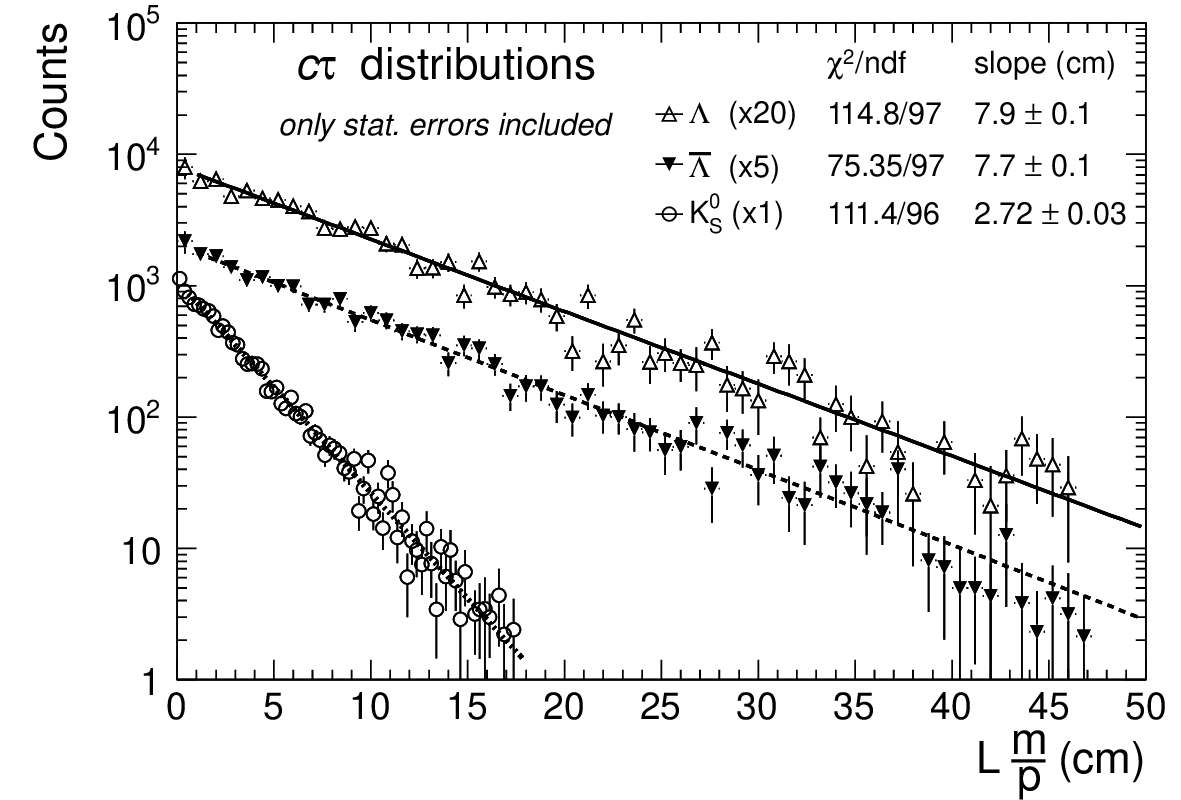}
}
\caption{$\Kzs$, $\rmLambda$ and $\rmAlambda$ lifetime distributions
  obtained for the 
candidates selected by the invariant mass within a $\pm 3\sigma$ region
  around the nominal mass and corrected for
detection efficiency.
The distributions are scaled for visibility and fitted to an exponential distribution (straight lines).
Only statistical uncertainties are shown.}
\label{fig:lifetime}
\end{figure}

The lifetime (c$\tau$) distributions for $\Kzs$, $\rmLambda$ and $\rmAlambda$ are
also checked.
All V0 candidates within  a $\pm 3\sigma$ effective mass region around the nominal
value are used in the distribution without further residual background subtraction.
The corresponding distributions of $c\tau=L\frac{m}{p}$ are obtained, where $L$ is
defined as the distance between primary and V0 vertices, and $m$ and $p$
are the particle mass and momentum.
Because of the acceptance, the single track efficiency and the topological selections
applied at reconstruction level, the reconstruction efficiency as a function of the decay
length is not constant.
The corresponding corrections are extracted from the reconstruction of full Monte Carlo
simulations (see section~\ref{subsec_eff}).
The corrected  $c\tau$ distributions are fitted using exponential functions. 
The results are shown with the statistical uncertainties in~Fig.~\ref{fig:lifetime}.
The extracted decay lengths of $7.9 \pm 0.1~\cm$, $7.7 \pm 0.1~\cm$ and 
$2.72 \pm 0.03~\cm$ for $\rmLambda$, $\rmAlambda$ and $\Kzs$, respectively, are
compatible with the PDG values given in Table~\ref{tab:3}.

\subsubsection{ $\phi$ reconstruction}
\label{sec_phireco}
The $\phi$ resonance is reconstructed through its principal decay channel
$\phi\rightarrow \rm{K}^+ \rm{K}^-$ (see  Table~\ref{tab:3}).
With a $c\tau$ of $45~\fm$, its decay vertex is indistinguishable from 
the primary collision vertex.
Therefore the selection criteria adopted for the candidate daughter tracks 
are the ones used for primaries, as specified in Table~\ref{tab:1}.

A crucial issue for the $\phi$ reconstruction, as for any strongly decaying
resonance, is the combinatorial background determination.
In the present analysis PID is used to select kaons, rejecting most
of the background while leading to a very small loss in efficiency.
For this purpose, tracks are selected
if the PID information from the TPC is compatible with a kaon signal and
using the TOF signal when available.

\begin{figure}
\resizebox{0.5\textwidth}{!}{%
  \includegraphics{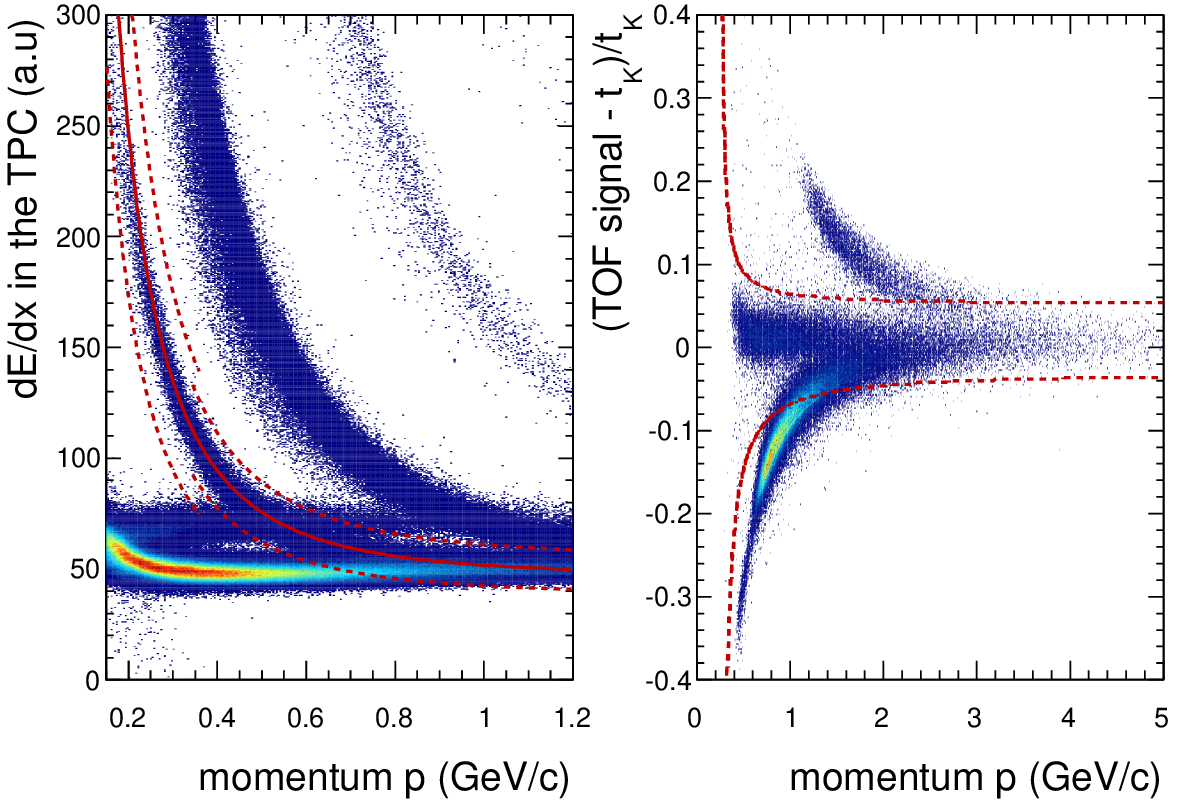}
}
\caption{Left panel shows the distribution of the measured energy-deposit of
charged particles versus their momentum in the TPC.
The dashed lines delimit successively a $\pm 5\sigma$ then a $\pm 3\sigma$
selection of the kaon tracks using the ALEPH parameterization~\cite{ALEPH} 
of the Bethe-Bloch curve (solid line).
Right panel shows the relative difference between TOF measured times and that 
corresponding to a kaon mass hypothesis.
The dashed lines delimit a coarse fiducial region compatible with this kaon hypothesis.}
\label{fig:dedxtof}
\end{figure}

For each track, the expected energy loss is calculated using a parametrised
response  based on the Bethe-Bloch formula~\cite{ALEPH} computed
with a kaon mass hypothesis.
It is compared with the TPC specific ionization $\dEdx$ measured via truncated
mean (the reconstructed momentum being evaluated at the  inner radius of the
TPC).
With the current TPC calibration for this data set, the assumed $\dEdx$ resolution
is $6~\%$.
For momenta smaller than $350~\mmom$, the species are well separated so the
window is set to $\pm 5\sigma$ with little or no contamination; above $350~\mmom$,
it is set instead to $\pm 3\sigma$ as shown in the left panel of Fig.~\ref{fig:dedxtof}.

The accepted band for TOF kaon identification is defined with two hyperbolas as
shown in the right panel of Fig.~\ref{fig:dedxtof}.
\subsection{Background evaluation and signal extraction}
\label{sec_bkgsig}
For minimum bias pp collisions, the signals for all particles are clearly distinguishable
from the combinatorial background as shown in Fig.~\ref{fig:invmass}.
Two different methods are used to extract the invariant mass signal
from the background.
For the single strange particles ($\Kzs$, $\rmLambda$ and $\rmAlambda$),
the signal is first approximated by a Gaussian on a second order polynomial
background.
This gives an estimate of the signal mean and width although the invariant
mass signal is not strictly Gaussian.
Then the background is sampled on each side of the signal by using
both sampled regions that are more than $6\sigma$ away from the Gaussian
mean.
The assumption that no reconstructed signal is included in these regions
is checked using Monte Carlo data.
The width of the background regions can vary depending on the $\pT$ interval
considered in the invariant mass distributions.
The sum of signal and background ($S$+$B$) is sampled in the region defined by
the mean $\pm 4\sigma$. 

\begin{figure}
\resizebox{0.5\textwidth}{!}{%
  \includegraphics{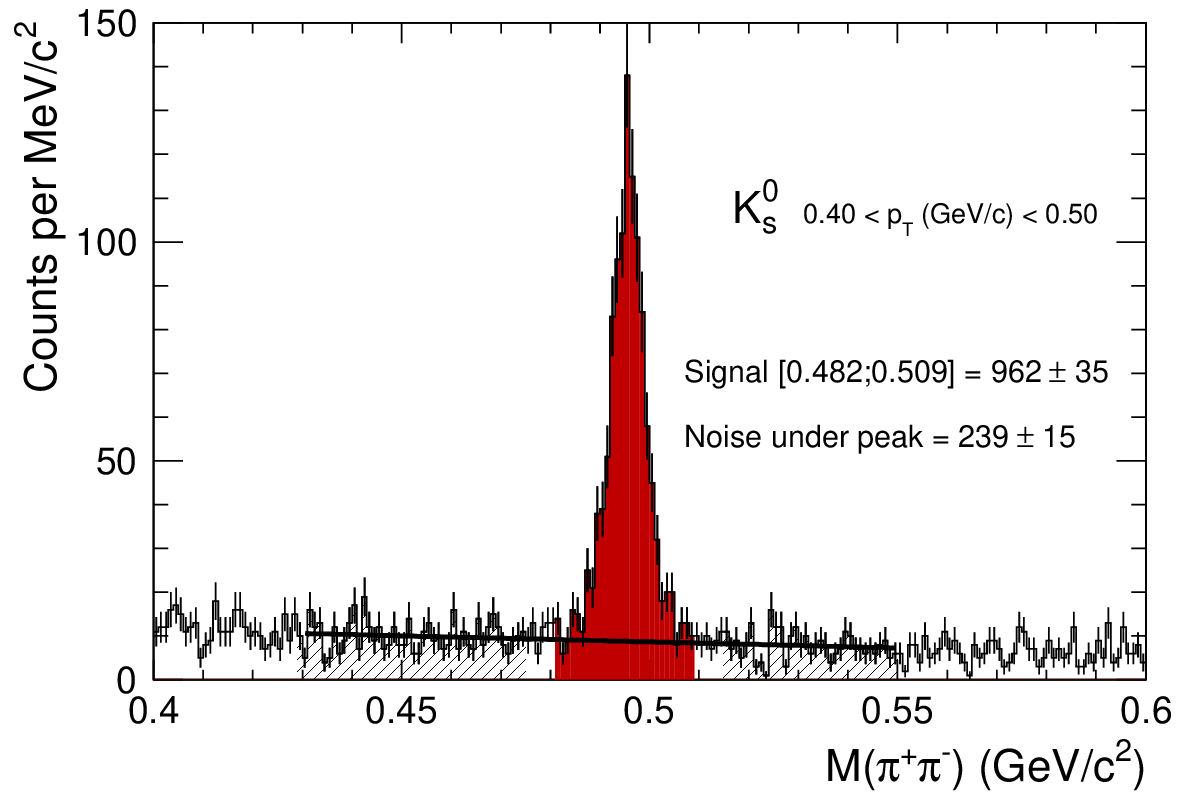}
}
\caption{Plot illustrating the ``bin-counting'' method used to extract the raw yields.
It corresponds to the invariant mass distribution of $\Kzs$ for the $\pT$ bin $[0.4-0.5]~\gmom$.
The hashed regions show where the background is sampled; they are chosen to be $6\sigma$
away from the signal approximated with a Gaussian distribution.
The averaged or fitted background is subtracted from the signal region of $\pm 4\sigma$.}
\label{fig:bincount}
\end{figure}

\begin{figure}
\resizebox{0.5\textwidth}{!}{%
  \includegraphics{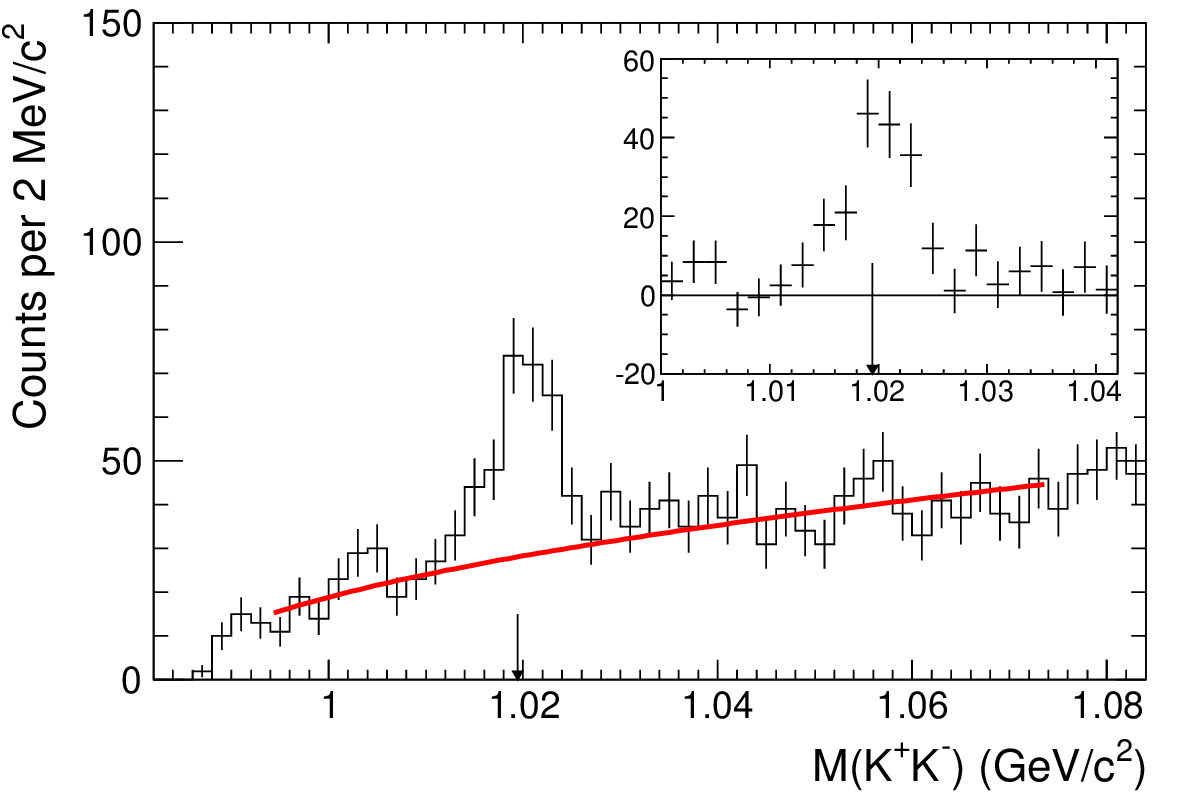}
}
\caption{Background evaluation for the $\phi$ corresponding to the $\pT$ bin [1.0-1.5]~$\gmom$.
The inset shows the $\phi$ signal after background subtraction.
The vertical arrows indicate the nominal mass value from PDG.}
\label{fig:phibkg}
\end{figure}

The sampling method is illustrated in Fig.~\ref{fig:bincount} for the $\Kzs$.
Two methods are used to evaluate the background and give consistent results.
The background areas are either i) fitted simultaneously with polynomial functions
(from first to third order) or ii) averaged by simply counting the number of entries 
(``bin-counting'').
The background $B$ under the signal $S$ is estimated using the normalized area
sampled on both sides of the signal region (Gaussian mean $\pm 4\sigma$).
The signal yield $S=(B+S)-B$ is thus evaluated without any assumption as to its shape.
Systematic effects such as signal asymmetry are taken into account by varying the
size of the signal and background intervals up to $1\sigma$.
The difference between the two methods (fit and bin-counting) contributes
to the evaluation of the systematic uncertainties associated to the signal extraction.
In the case of the $\Xis$, statistical uncertainties are significant so that, in parallel to the
bin-counting method, the background level is simply estimated by a straight line fit.
The $\phi$ invariant mass distribution has a larger combinatorial background
and a function reproducing both the background and the signal is preferred.
It is found that the background can be well reproduced by a function f(M) = $a\sqrt{M-b}$,
while the peak has the shape of a Gaussian. 
The peak range is defined as $\pm 4\sigma$ around the PDG mass of the  $\phi$,
where $\sigma = \Gamma/2.35$ and $\Gamma$ is the nominal value of the resonance
full width at half maximum ($4.5~\mmass$)~\cite{Nakamura:2010zzi}.
For each analyzed $\pT$ bin, several fit ranges are investigated.
It is found that the fitted width matches that extracted from a full Monte Carlo
simulation (as defined in section~\ref{subsec_eff}) within $5\%$, except for
the last $\pT$ bin where it is broader ($\sim 10\%$).
While fluctuations of the fit values as a function of the fit range are taken
into account for the systematic error (see section~\ref{sec:systselsig}),
the fit values used for all subsequent steps in the analysis are those that
minimize the difference $| \chi^{2}/\rm{NDF} - 1|$.
Figure~\ref{fig:phibkg} illustrates the method for the $[1.0-1.5]~\gmom$ $\pT$ bin.
Every unlike-sign track pair passing all selection criteria and falling within the
$\phi$ invariant mass peak range is counted.
The total number of $\phi$ is estimated by subtracting the integral of the
background function alone, computed in the same invariant mass range.
\bigskip

\begin{figure}
\resizebox{0.5\textwidth}{!}{%
  \includegraphics{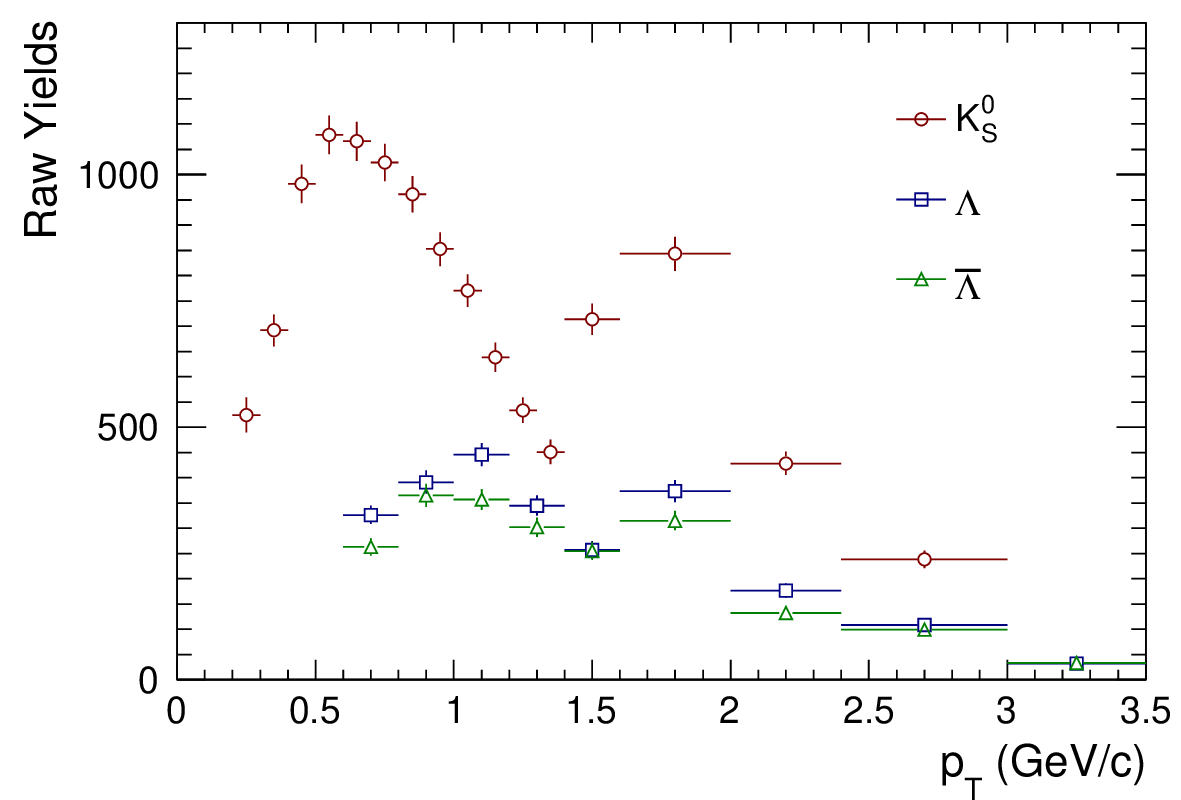}
}
\caption{Reconstructed (raw) yields of $\Kzs$ (open circles), $\rmLambda$ (open squares)
and $\rmAlambda$ (open triangles) as a function of $\pT$. The change of bin size results
in successive offsets of the raw counts at $\pT~=~[1.4, 1.6, 2.4] ~\gmom$ for $\Kzs$ and 
$\pT~=~[1.6, 2.4, 3.0] ~\gmom$ for $\rmLambda$ and $\rmAlambda$.
Uncertainties correspond to the statistics and the systematics from the signal extraction.
They are represented by the vertical error bars. The horizontal error bars give the bin width.}
\label{fig:rawv0}
\end{figure}

\begin{figure}
\resizebox{0.5\textwidth}{!}{%
  \includegraphics{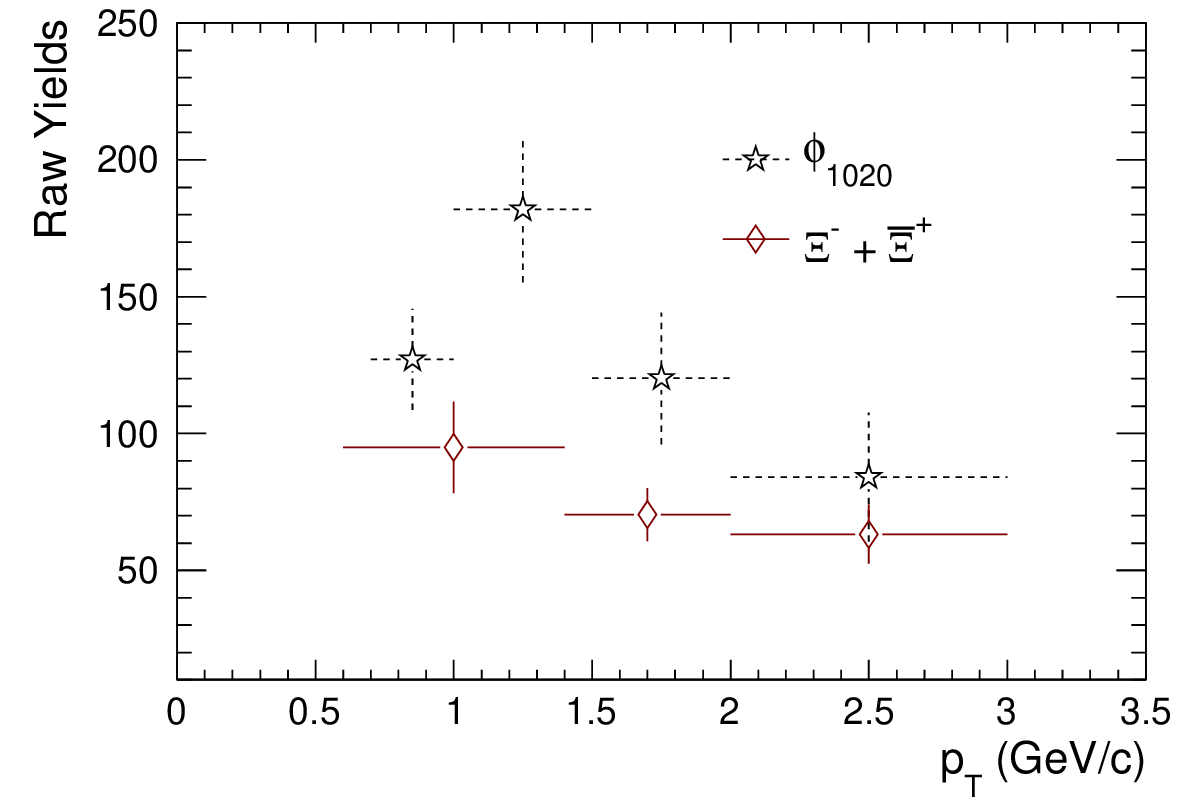}
}
\caption{Reconstructed (raw) yields of $\phi$ (stars) and $\Xis$ (diamonds) as a function of $\pT$.
With the current statistics, $4~\pT$ bins are used for the $\phi$ ([0.7-1.0], [0.7-1.5], [1.5-2.0]
and [2.0-3.0]~$\gmom$)
and $3~\pT$ bins for the $\Xis$ ([0.6-1.4], [1.4-2.0] and [2.0-3.0]~$\gmom$).
Uncertainties correspond to the statistics (i.e. the number of reconstructed particles)
and the systematics from the signal extraction. They are represented by the vertical error bars
(the horizontal ones give the bin width).}
\label{fig:rawxp}
\end{figure}

The signal counts (raw yields) for each of the $\pT$ bins are histogrammed as a function
of $\pT$ for $\Kzs$, $\rmLambda$, $\rmAlambda$ in Fig.~\ref{fig:rawv0} and for
$\phi$ and $\Xis$ in Fig.~\ref{fig:rawxp}.
The uncertainties correspond to both the statistical errors related to the
number of counts and the systematics from the bin-counting and fit methods used to
extract the signal from the background. 
\subsection{Efficiency corrections}
\label{subsec_eff}
\begin{figure}
\resizebox{0.5\textwidth}{!}{%
  \includegraphics{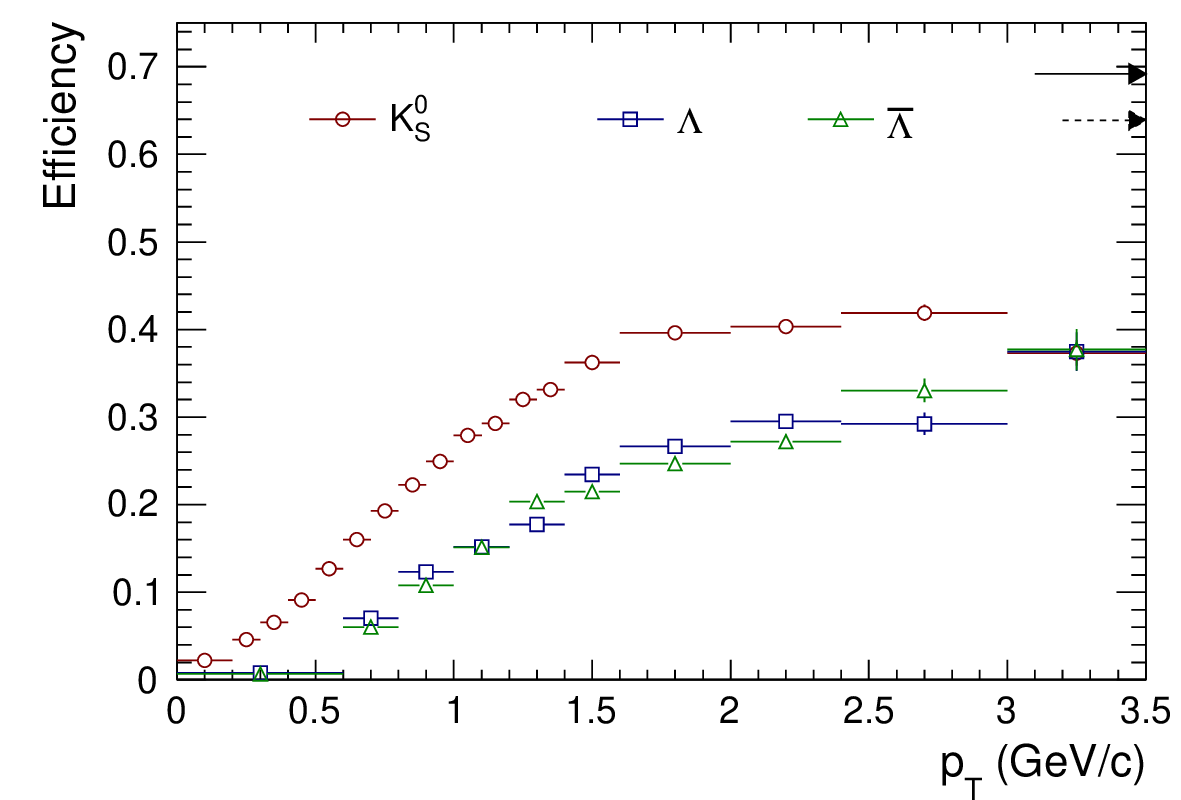}
}
\caption{Efficiency of $\Kzs$ (open circles), $\rmLambda$ (open squares) and $\rmAlambda$
(open triangles) as a function of $\pT$.
The uncertainties correspond to the statistics in Monte Carlo samples used to compute the
corrections.
The efficiency is limited by the branching ratio represented by a solid arrow for $\Kzs$ (0.692)
and by a dashed arrow for $\rmLambda$ and $\rmAlambda$ (0.639).
}
\label{fig:effv0}
\end{figure}

\begin{figure}
\resizebox{0.5\textwidth}{!}{%
  \includegraphics{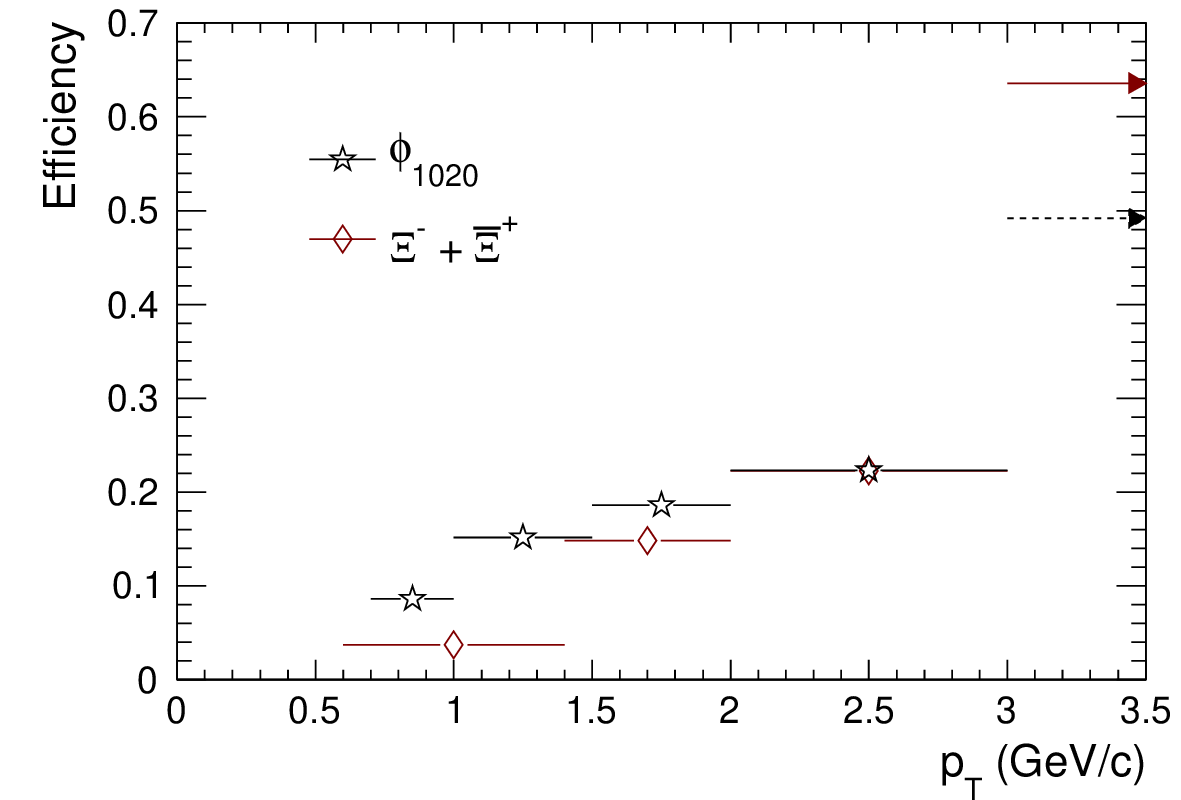}
}
\caption{Efficiency of $\phi$ (stars) and $\Xis$ (diamonds)
as a function of $\pT$. The uncertainties correspond to the statistics
in the Monte Carlo sample used to compute the corrections.
The efficiency is limited by the branching ratio represented by a solid arrow for $\Xis$ (0.636)
and by a dashed arrow for $\phi$ (0.492).
}
\label{fig:effxp}
\end{figure}

\begin{figure}
\resizebox{0.5\textwidth}{!}{%
  \includegraphics{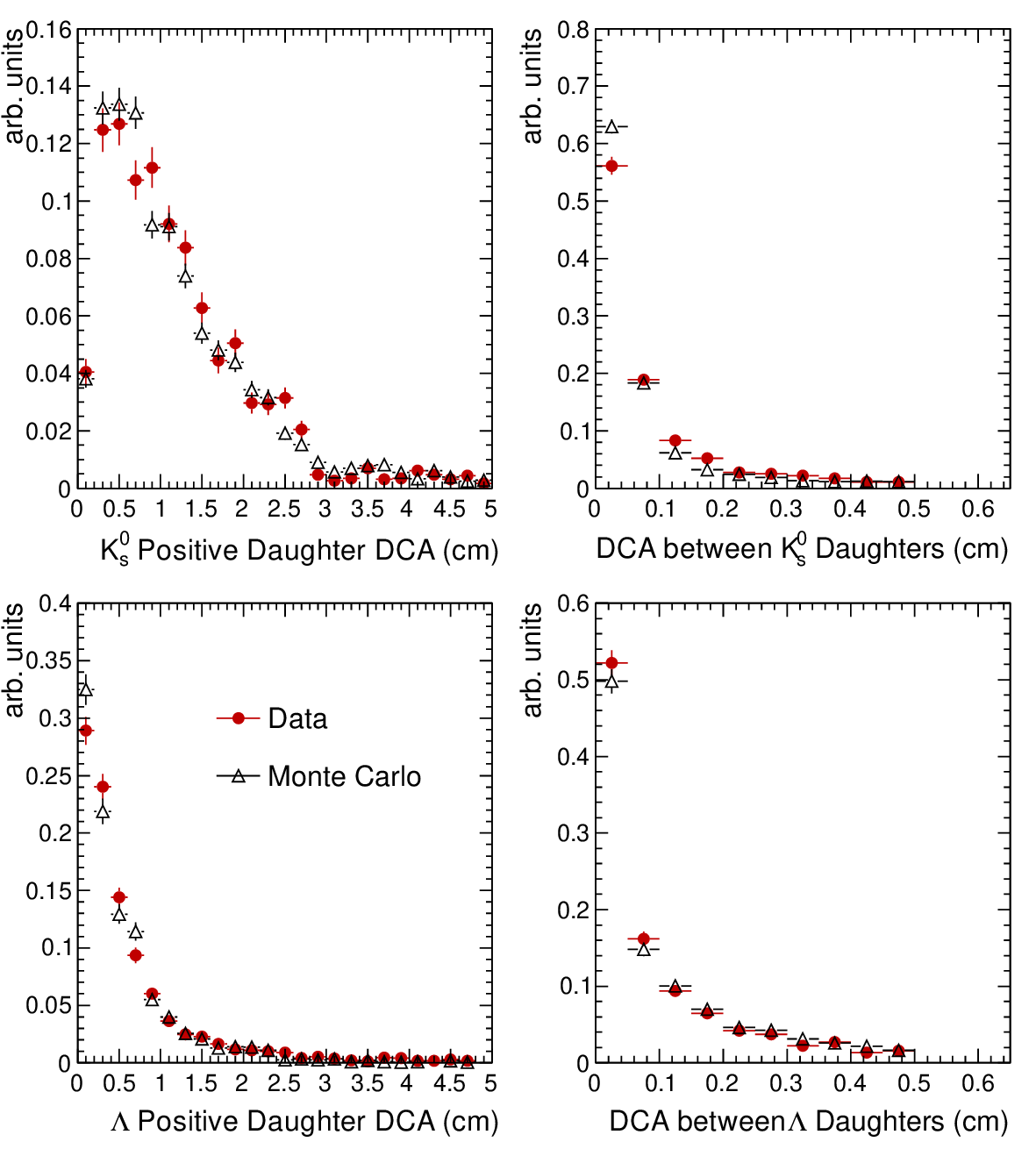}
}
\caption{Comparison between data (red circles) and Monte Carlo (black open triangles)
for several topological variables used to select secondary vertices. 
The top panels correspond to the $\Kzs$ candidates selected in a $\pm 20~\mmass$
invariant mass window around the nominal mass.
The distribution of the DCA between the positive daughter track and the primary vertex
and the DCA distribution between the two daughters tracks are displayed in the left and
the right top panels respectively.
On the bottom panels, the same distributions are shown for the $\rmLambda$ candidates
selected in a $\pm 8~\mmass$ invariant mass window around the nominal mass.}
\label{fig:selv0}
\end{figure}

The efficiency corrections are obtained by analysing Monte Carlo (MC) events in exactly
the same way as for the real events.
Little dependence is found on the several MC generators which are used.
Therefore the corrections presented here are obtained using
the event generator PYTHIA 6.4 (tune D6T)~\cite{Sjostrand:2006za,Albrow:2006rt} and 
GEANT3~\cite{Alice:Geant} for particle transport through the ALICE detectors.
The MC information is propagated through the whole reconstruction and identification
procedure to generate the differential $\pT$ efficiencies as shown in Fig.~\ref{fig:effv0}
for $\Kzs$, $\rmLambda$ and $\rmAlambda$ and in Fig.~\ref{fig:effxp} for $\phi$ and $\Xis$.
The uncertainties correspond to the statistics of Monte Carlo samples used to compute the
corrections. 
For all particles, the global efficiency is limited at low $\pT$ because of the acceptance of
at least two charged daughter tracks in the detection volume of the TPC (three tracks in
the case of $\Xis$).
It rapidly increases with $\pT$ but cannot exceed the asymptotic limits given by the
charged particle decay branching ratios presented in Table~\ref{tab:3}.
The difference between the $\rmLambda$ and $\rmAlambda$ reflects the absorption
of the anti-proton daughter of the $\rmAlambda$.
For all the variables used to select the particles and improve the signal
over noise ratio (see Tables~\ref{tab:1} and~\ref{tab:2}), it is verified that data
and MC distributions match, thus possible efficiency biases can be properly managed.
Examples of such distributions are presented in Fig.~\ref{fig:selv0}.

\subsection{Estimation of the systematic uncertainties}
\label{subsec_sys}
Systematic uncertainties are discussed in the following sections, where
details are given on the contributions due to topological selections and
signal extraction methods, as well as those due to material budget and
feed-down.
As for efficiency corrections, MC data are generated with PYTHIA 6.4 
(tune D6T)~\cite{Sjostrand:2006za,Albrow:2006rt} and transported with
GEANT3~\cite{Alice:Geant}.
At low $\pT$, the anti-proton absorption cross section in GEANT3 is known
to be too large~\cite{Aamodt:2010dx,Alice:AbsAIN}.
GEANT4 (with the absorption cross-sections of~\cite{Bendiscioli:1994uv})
was then used to correct the anti-proton tracking efficiency.
The information is summarized in Table~\ref{tab:4}.
In addition to these point-to-point systematic uncertainties, there is also a $2~\%$
systematic error on the global normalization coming from the evaluation of the total
number of inelastic events.
%
%
\begin{table*}
\caption{Point-to-point systematic uncertainties expressed in percentage for $\pT$ spectra of different particles.
For each particle, the reported values correspond to the effect on the lowest $\pT$ bin, the average and the
highest $\pT$ bin, except for the feed-down contributions where values are estimated as being constant versus
$\pT$ or where the effect is found to be negligible (less than 2 standard deviations from the default value on the
corrected spectrum).}
\label{tab:4}       
\begin{center}
\resizebox{1.00\textwidth}{!}{
\begin{tabular}{lcccccc}
\hline\noalign{\smallskip}
\multicolumn{2}{c}{systematic effects ($\%$)} &         $\Kzs$          &   $\rmLambda$     &  $\rmAlambda$   &          $\phi$            &         $\Xis$             \\
\noalign{\smallskip}\hline\noalign{\smallskip}
\multicolumn{2}{l}{Selections}                  &    \multicolumn{5}{c}{ }  \\
                                            & tracks             & $[4.6-1.1-2.1]$ & $[2.6-2.0-2.5]$   & $[3.0-2.0-4.1]$  &  $[0.9-3.1-6.0]$  & $[negl.-5.4-negl.]$   \\
                                            & topological    & $[3.8-1.4-1.3]$ & $[3.3-3.3-1.5]$   & $[4.7-4.7-3.8]$  &  $--$  & $[6.8-11.6-13.9]$   \\
                                            \noalign{\smallskip}\hline\noalign{\smallskip}
\multicolumn{2}{l}{Signal extraction}  &             $[4.5-1.5-1.5]$           & $[3.0-2.0-5.0]$       &  $[3.0-2.0-5.0]$       & $[3.2-4.3-7.0]$              & $[5.6-negl.-2.5]$   \\   
                                            \noalign{\smallskip}\hline\noalign{\smallskip}
\multicolumn{2}{l}{TPC $\dEdx$} & $--$ & $[5-negl.]$   & $[5-negl.]$  &  $[1.8-2.9-3.6]$  & $[negl.]$   \\
                                            \noalign{\smallskip}\hline\noalign{\smallskip}
\multicolumn{2}{l}{Efficiency}                    &    \multicolumn{5}{c}{ }  \\
                                       & material budget & $[1.5-1.5-1.1]$ & $[3.4-1.0-1.6]$   & $[3.7-2.0-4.5]$  &  $[4.7 - 4.0 - 2.3]$  & $[2.7-1.5-3.6]$   \\
                          & $\bar{p}$ cross-section & $--$ & $<1$   & $<2$  &  $--$  & $<2$   \\
                                            \noalign{\smallskip}\hline\noalign{\smallskip}
\multicolumn{2}{l}{Feed-down}                &             $--$           & $1.7$   & $1$  &  $--$  & $--$   \\   

\noalign{\smallskip}\hline
\end{tabular}
}
\end{center}
\end{table*}

\subsubsection{Systematic uncertainties due to track or topological selections and signal extraction}
\label{sec:systselsig}
Systematic uncertainties due to tracking and topological identification are determined
by varying the track and topological (for secondary vertices) selections, as well as the
definition of the regions sampled for signal extraction.
To assess the different systematic uncertainties, only the deviations that are statistically
significant are taken into account (more than 2 standard deviations away from the central 
value on the corrected spectrum).

The systematic variation of track and topological selections results in a variation of the
amount of signal extracted from invariant mass distribution in both data and the Monte
Carlo simulation mentioned above.
The difference between these amounts of signal corresponds indirectly to the accuracy
with which the MC simulation reproduces the characteristics of real events, from the 
simulation of the detector response to the background shape and composition
considered for the extracted signal.
It is estimated that the point-to-point uncertainties in the $\pT$ spectra are at most
$4.6~\%$, $3.3~\%$, $4.7~\%$, $6~\%$ and $13.9~\%$ for the $\Kzs$, $\rmLambda$,
$\rmAlambda$, $\phi$ and $\Xis$, respectively.
The systematic uncertainties of the signal extraction for $\Kzs$, $\rmLambda$,
$\rmAlambda$ and $\Xis$ are $\pT$-dependent and estimated by varying the invariant
mass regions where the signal and the background are sampled using the bin-counting
method described in section~\ref{sec_bkgsig}.
For the $\phi$ signal, the systematic uncertainties from background subtraction are
estimated using three different criteria.
First of all, the function reproducing the background is replaced by a second or third
order polynomial.
Moreover, the fit is repeated fixing the width parameter to $\pm 10 \%$ of the value
obtained in the default procedure (described in section \ref{sec_bkgsig}), and also
to the value obtained when fitting the Monte Carlo sample and to $\pm 10 \%$ of this.
Finally, the fit range is also varied.
All of these computations result in a variation of the raw counts with respect to those
shown in Fig.~\ref{fig:rawxp}.
Although a quite large compatibility region is requested for PID (at least $3\sigma$)
the effects of varying the $\dEdx$  selections are taken into account for the corresponding
efficiency calculation.
For both $\phi$ and $\Xis$, statistical errors dominate after signal extraction
(see section~\ref{sec_bkgsig}) and consequently, some systematic effects
due to PID are extrapolated from single track and V0 measurements.
The TOF PID selection is applied only to reject the $\phi$ background.
No systematic effects are observed on the $\phi$ signal i) for the Monte Carlo
data sample, when the selection is applied to the $\phi$ daughters in addition
to all other cuts; ii) for real events, when comparing the $\phi$ statistics
before and after applying the selection.
\subsubsection{Systematic uncertainties due to material budget and absorption cross-section}
\label{sec:systmatabs}
A dedicated study involved the variation of the detector material thickness crossed
by particles.
The material budget uncertainty, based on $\gamma$ conversion measurements,
is estimated to be $7\%$ in terms or radiation length~\cite{Aamodt:2010dx}.
The efficiency variation due to this material budget uncertainty depends on the
momentum of each of the decay daughters.
Although such a variation is also correlated with the momentum of the parent particle,
the corresponding systematic uncertainties are reported as point-to-point errors in
Table~\ref{tab:4} for the lowest, the average and the highest $\pT$ bin and eventually
added in quadrature to the total systematic errors.

Specific uncertainties are related to the (anti-)proton absorption and scattering
cross-sections used for propagating these particles through the geometry of the
detectors with both GEANT3~\cite{Alice:Geant} (and its default absorption
cross-sections) and GEANT4 (using the absorption cross-sections of~\cite{Bendiscioli:1994uv}).
More details about the modifications can be found in~\cite{Aamodt:2010dx,Alice:AbsAIN}
and references therein.
The corresponding corrections are taken into account in the efficiency versus $\pT$
assuming that absorption cross-sections are identical for the (anti-)~hyperon and its 
(anti-)~proton daughter.
The uncertainties associated with these corrections are derived from the (anti-)proton
cross-section uncertainties and the values are estimated as constant and lower than
$1\%$ ($2\%$) for $\rmLambda$ ($\rmAlambda$) and  $2\%$ for $\Xis$.
\subsubsection{Systematic uncertainties for $\rmLambda$ and $\rmAlambda$ due to feed-down}
\label{sec:sysfddown}
Some of the reconstructed $\rmLambda$ and $\rmAlambda$ particles come
from decays of $\mathrm{\Xi}$-hyperons.
The proportion of reconstructed secondary $\rmLambda$ and $\rmAlambda$
depends on the selection criteria used.
For the parameters listed in Table~\ref{tab:2} (V0 vertex part), the impact of the $\mathrm{\Xi}$ feed-down
on the final spectra is evaluated to be $13\%$ for $\rmLambda$ and $12\%$ for
$\rmAlambda$.
No $\pT$ dependence is found within uncertainties.

This assessment results in a global correction of the spectra, applied as an additional
factor in the overall normalization.
Provided that both primary and secondary $\rmLambda$ have similar spectral shapes,
such integrated correction is applicable.
This is tested directly using Monte Carlo data, but also with real data, changing the fraction
of the secondary $\rmLambda$ by varying the DCA of reconstructed candidates. 
Within the available statistics and $\pT$ reach, no significant change in spectral
shape is observed.

Using Monte Carlo, the ratio $r_{\rm{feed-down}}$ of the reconstructed $\rmXi$
($\rmAxi$) candidates to the number of reconstructed $\rmLambda$ ($\rmAlambda$) 
candidates from $\mathrm{\Xi}$ decays is:
$$ r_{\rm{feed-down}} = \frac{(N_{\rmXi})_{\rm{MC}}}{(N_{\rmLambda \leftarrow \rmXi})_{\rm{MC}}} $$
Assuming that this ratio is the same in both Monte Carlo and data, the whole feed-down
contribution to the spectra is estimated by dividing the number of reconstructed
$\rmXi$ ($\rmAxi$) in data by the ratio extracted from Monte Carlo:
$$ (N_{\rmLambda \leftarrow \rmXi})_{\rm{data}} = \frac{(N_{\rmXi})_{\rm{data}}}{r_{\rm{feed-down}}} $$
Besides the $\mathrm{\Xi}$ contribution, other sources may feed the $\rmLambda$
population resulting in additional systematic uncertainties.
In Monte Carlo simulations, $\rmLambda$ particles possibly generated in the detector
material induce a $1.7~\%$ uncertainty.
The same uncertainty in the case of $\rmAlambda$ is below $1~\%$.
The contribution from $\mathrm{\Omega}$ decays is found to be negligible.
It should be noted that since $\rmLambda$ ($\rmAlambda$) from electromagnetic $\rm{\Sigma}^{0}$
($\rm{\overline{\Sigma}}^{0}$) decays cannot be distinguished from the direct ones, the identified
$\rmLambda$ ($\rmAlambda$) include these contributions.
\subsection{$\pT$ spectra and global yield extraction}
\label{subsec_sgy}
\begin{figure}
\resizebox{0.5\textwidth}{!}{%
  \includegraphics{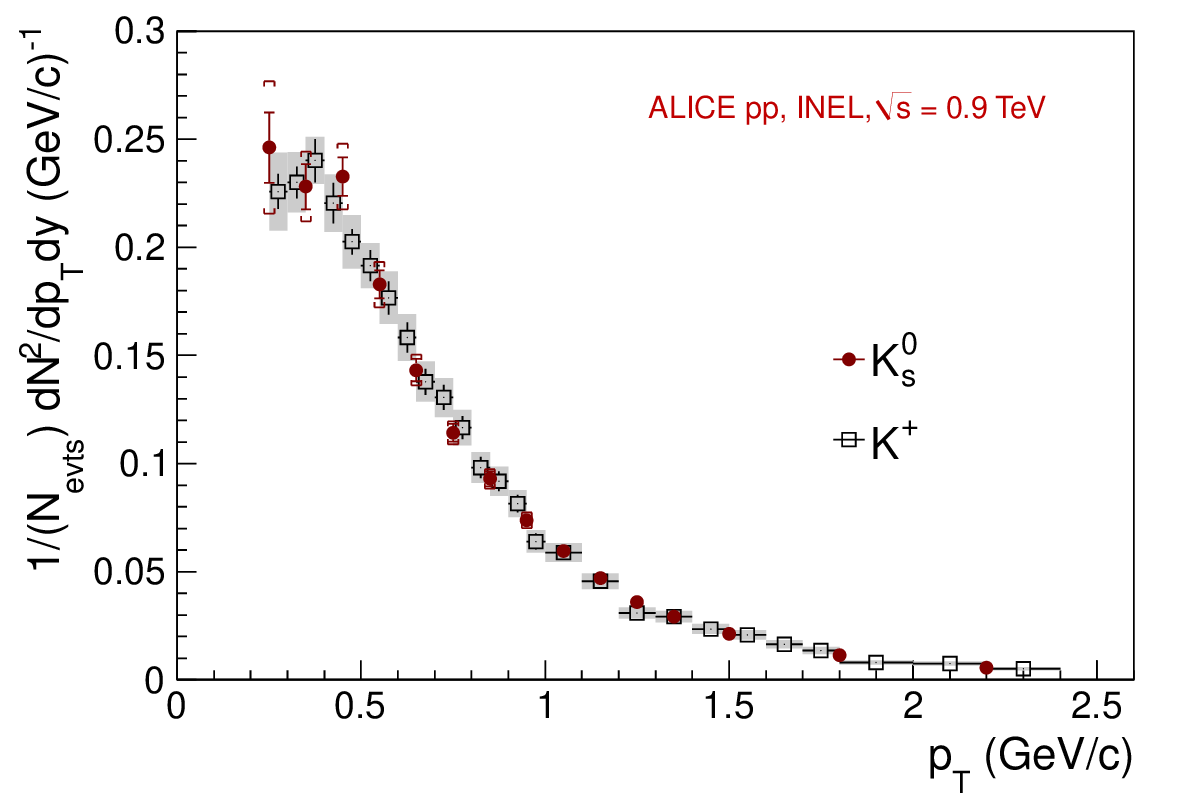}
}
\caption{Comparison of the corrected yields as a function of $\pT$ for $\Kzs$ (circle) and charged
kaons ($\mathrm{K^{+}}$) (open squares), identified via energy loss in the TPC and ITS, and via 
time of flight in the TOF.
The points are plotted at the centre of the bins.
The full vertical lines associated to the $\Kzs$ points, as well as the gray shaded areas associated
to the $\mathrm{K^{+}}$ points, correspond to the statistical and systematic uncertainties summed
in quadrature whereas the inner vertical lines contain only the statistical uncertainties (i.e. the
number of reconstructed particles) and the systematics from the signal extraction.}
\label{fig:compkaon}
\end{figure}
The $\Kzs$ spectrum is first shown on a linear scale in Fig.~\ref{fig:compkaon} and compared
with charged kaon spectra~\cite{Alice:Pid}.
Within uncertainties, good agreement is found between $\Kzs$ and $\rm{K}^{+}$
in the measured $\pT$ range.

Figure~\ref{fig:corryield} presents the corrected  $\pT$ spectra for all species, including both
statistical errors and systematic uncertainties.
The spectra are fitted with two different functional forms in order to extract the global
integrated yields: 
\begin{eqnarray}
\frac{d^{2}N}{dyd\pT}  &  =  & A \times \pT \times e^{- \frac{\pT}{T}} \label{eqn:funcpexp} \\
\frac{d^{2}N}{dyd\pT}  &  =  & \frac{(n-1)(n-2)} {nT [nT + m(n-2)]} \times \frac{dN}{dy} \nonumber \\
                                                               &       &\times \pT \times \left( 1+ \frac{\mT-m}{nT}\right)^{-n} \label{eqn:funclevy}
\end{eqnarray}
where $\mT = \sqrt{m^{2} + \pT^{2}}$.
The $\pT$ exponential has two parameters: the normalization $A$ and the inverse slope parameter $T$.
The L\'{e}vy function [Eq.~(\ref{eqn:funclevy})], already used at lower energies~\cite{Abelev:2006cs},
is shown to be useful when the $\pT$ range is wide:
it includes both an exponential shape for low $\pT$ (which can be characterized by an
inverse slope parameter $T$) and a power law component (governed by the power
parameter $n$) for the higher $\pT$ region.
The results of these fits to the spectra, where statistical and systematic uncertainties are
added in quadrature, are shown in Fig.~\ref{fig:corryield} and in Table~\ref{tab:5}.
In the case of $\Kzs$ for which the statistics and the $\pT$ range are larger than for other
species, the $\chi^{2}/\mathrm{NDF}$ indicates clearly that the $\pT$ exponential
parameterization cannot properly reproduce the spectrum shape.

For the spectra of the $\phi$, $\rmLambda$ and $\rmAlambda$ both functions give
similar and acceptable $\chi^{2}/\mathrm{NDF}$.
Within uncertainties, $\rmLambda$ and $\rmAlambda$ have the same fit parameters.
In the case of the $\Xis$ spectrum, the low number (i.e. $3$) of $\pT$ bins cannot constrain the
L\'{e}vy function and therefore its $\chi^{2}/\mathrm{NDF}$ in Table~\ref{tab:5} is not defined.
Nevertheless, for consistency and in order to extract particle ratios, a L\'{e}vy fit
is performed to obtain the integrated yields and particle ratios for all species.
It must be noted that the rapidity range is slightly different for each species (cf. Table~\ref{tab:6}).
However, the rapidity dependence of particle production at mid-rapidity is weak enough
to allow direct comparisons of the spectra~\cite{Aamodt:2010dx}.
%
%
\begin{table*}
\caption{Summary of the parameters extracted from the fits to the measured transverse momenta spectra using $\pT$ exponential (\ref{eqn:funcpexp}) and L\'{e}vy (\ref{eqn:funclevy}) functional forms and including point-to-point systematic uncertainties.
}
\label{tab:5}       
\begin{center}
\begin{tabular}{lcccccc}
\hline\noalign{\smallskip}
\multicolumn{2}{c}{}						&	\multicolumn{2}{c}{$\pT$ exponential (\ref{eqn:funcpexp})}	&	\multicolumn{3}{c}{L\'{e}vy (\ref{eqn:funclevy})} \\ 
\multicolumn{2}{c}{Particles} 	               & $T~(\mev)$     & $\chi^{2}$/NDF   & $T~(\mev)$         & $n$	                & $\chi^{2}$/NDF	\\ 
\noalign{\smallskip}\hline\noalign{\smallskip}
\multirow{2}{*}{Mesons}	        & $\Kzs$       & $325 \pm   4$	& $117.6/14$ 	   & $168 \pm  5$       & $ 6.6 \pm 0.3$        & $10.8/13$     \\ 
                                & $\phi$       & $438 \pm  31$	& $  1.3/ 2$ 	   & $164 \pm 91$       & $ 4.2 \pm 2.5$        & $ 0.6/ 1$     \\ 
					\noalign{\smallskip}\hline\noalign{\smallskip}                                           
\multirow{3}{*}{Baryons}  	& $\rmLambda$  & $392 \pm   6$	& $  10.2/7$	   & $229 \pm 15$       & $10.8 \pm 2.0$        & $9.6 / 6$     \\ 
                                & $\rmAlambda$ & $385 \pm   6$	& $  5.1/7$ 	   & $210 \pm 15$       & $ 9.2 \pm 1.4$        & $3.7 / 6$     \\ 
                                & $\Xis$       & $421 \pm  42$	& $  2.0/1$        & $175 \pm 50$       & $ 5.2 \pm 2.3$        & $--$          \\
\noalign{\smallskip}\hline 
\end{tabular}
\end{center}
\end{table*}

\section{Results and discussion}
\label{sec_res}%
\begin{figure}
\resizebox{0.5\textwidth}{!}{%
  \includegraphics{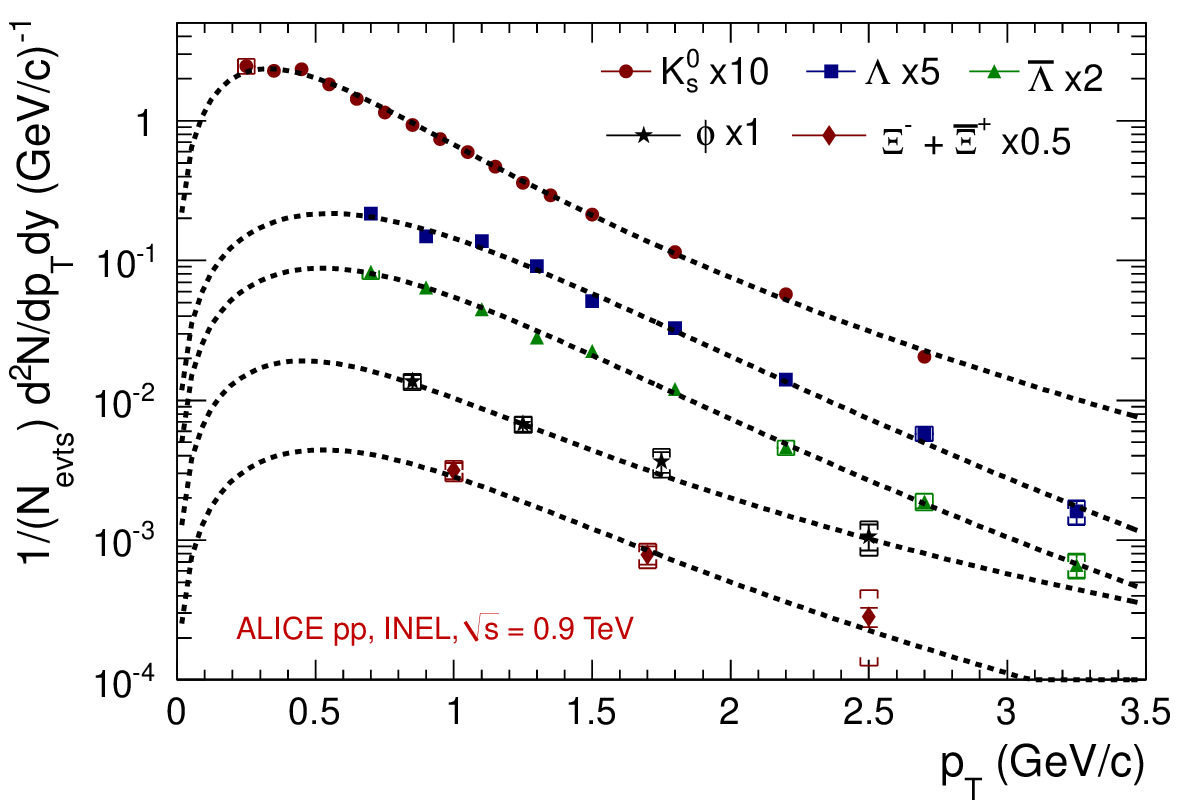}
}
\caption{Particle spectra (corrected yields) as a function of $\pT$ for $\Kzs$ (circles),
$\rmLambda$ (squares), $\rmAlambda$ (triangles), $\phi$ (stars) and $\Xis$
  (diamonds). The data points are scaled for visibility
and plotted at the centre of the bins.
Uncertainties corresponding to both statistics (i.e. the number of reconstructed particles)
and systematics from the signal extraction are shown as vertical error bars.
Statistical uncertainties and systematics (summarized in Table~\ref{tab:4}) added in
quadrature are shown as brackets.
The fits (dotted curves) using L\'{e}vy functional form [see Eq.~(\ref{eqn:funclevy})] are
superimposed.
}
\label{fig:corryield}
\end{figure}

\begin{figure}
\resizebox{0.5\textwidth}{!}{%
  \includegraphics{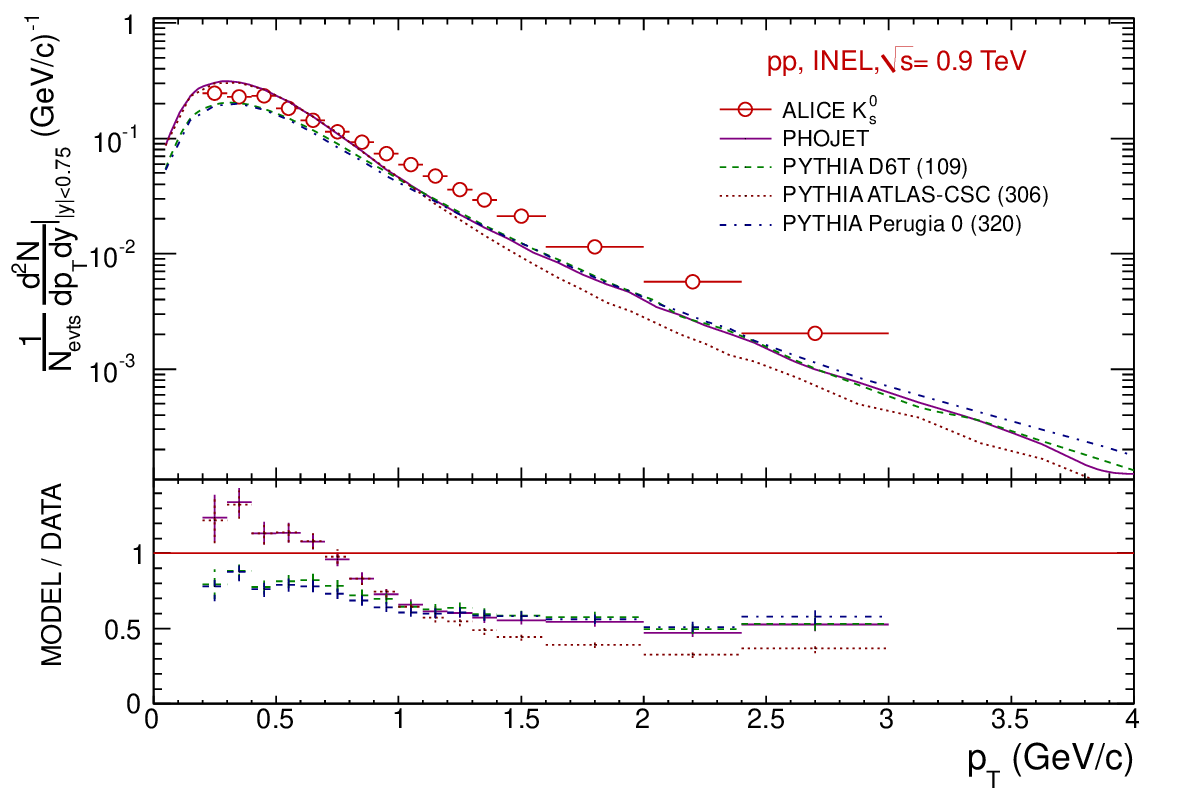}
}
\caption{Comparison of the transverse momentum differential yield for the $\Kzs$
particles for INEL pp collisions with PHOJET and PYTHIA tunes 109, 306 and 320.}
\label{fig:k0smc}
\end{figure}

\begin{figure}
\resizebox{0.5\textwidth}{!}{%
  \includegraphics{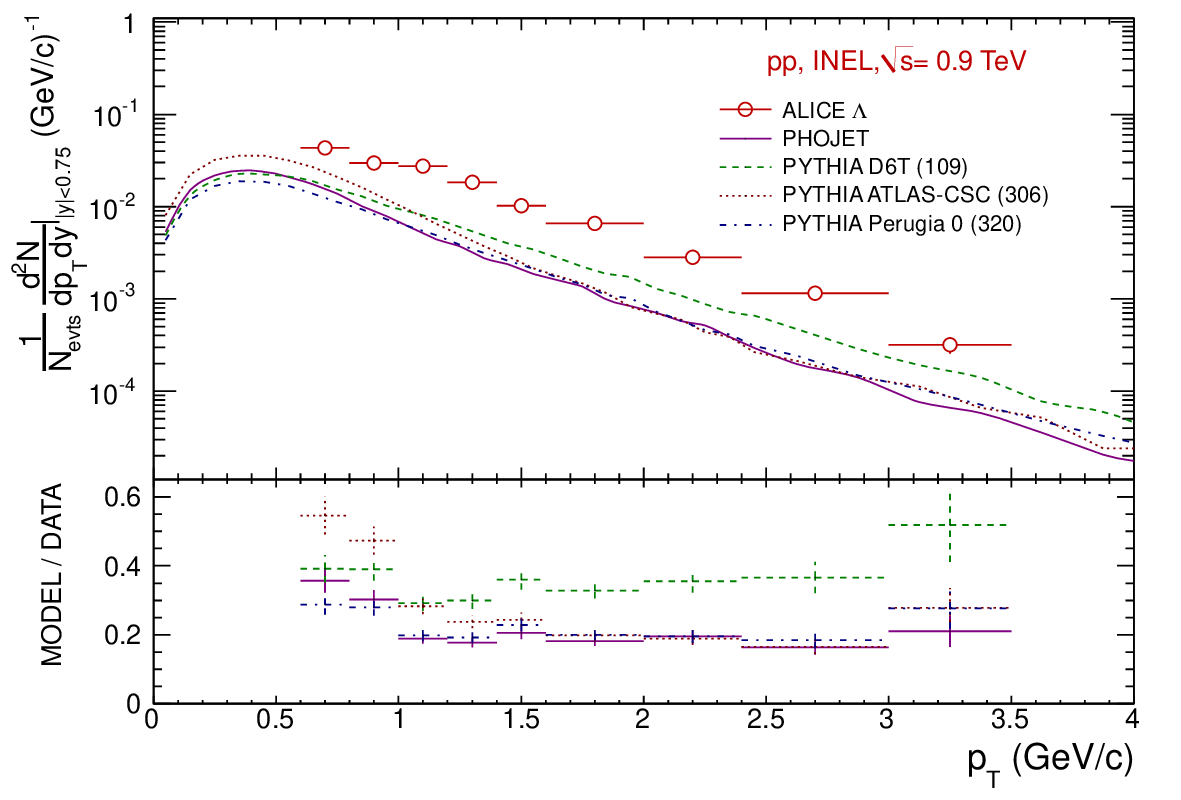}
}
\caption{Comparison of the transverse momentum differential yield for the $\rmLambda$
particles for INEL pp collisions with PHOJET and PYTHIA tunes 109, 306 and 320.
}
\label{fig:lammc}
\end{figure}

\begin{figure}
\resizebox{0.5\textwidth}{!}{%
  \includegraphics{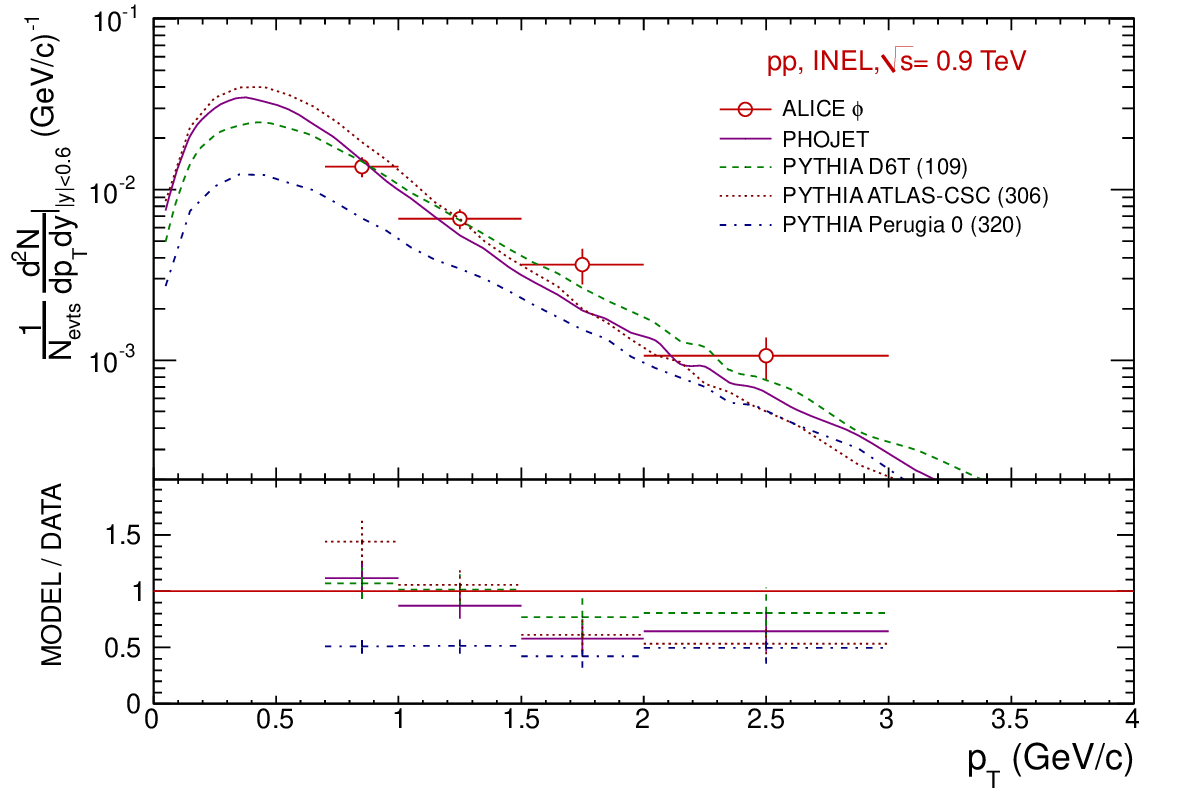}
}
\caption{Comparison of the transverse momentum differential yield for the $\phi$
particle for INEL pp collisions with PHOJET and PYTHIA tunes 109, 306 and 320.
}
\label{fig:phimc}
\end{figure}

\begin{figure}
\resizebox{0.5\textwidth}{!}{%
  \includegraphics{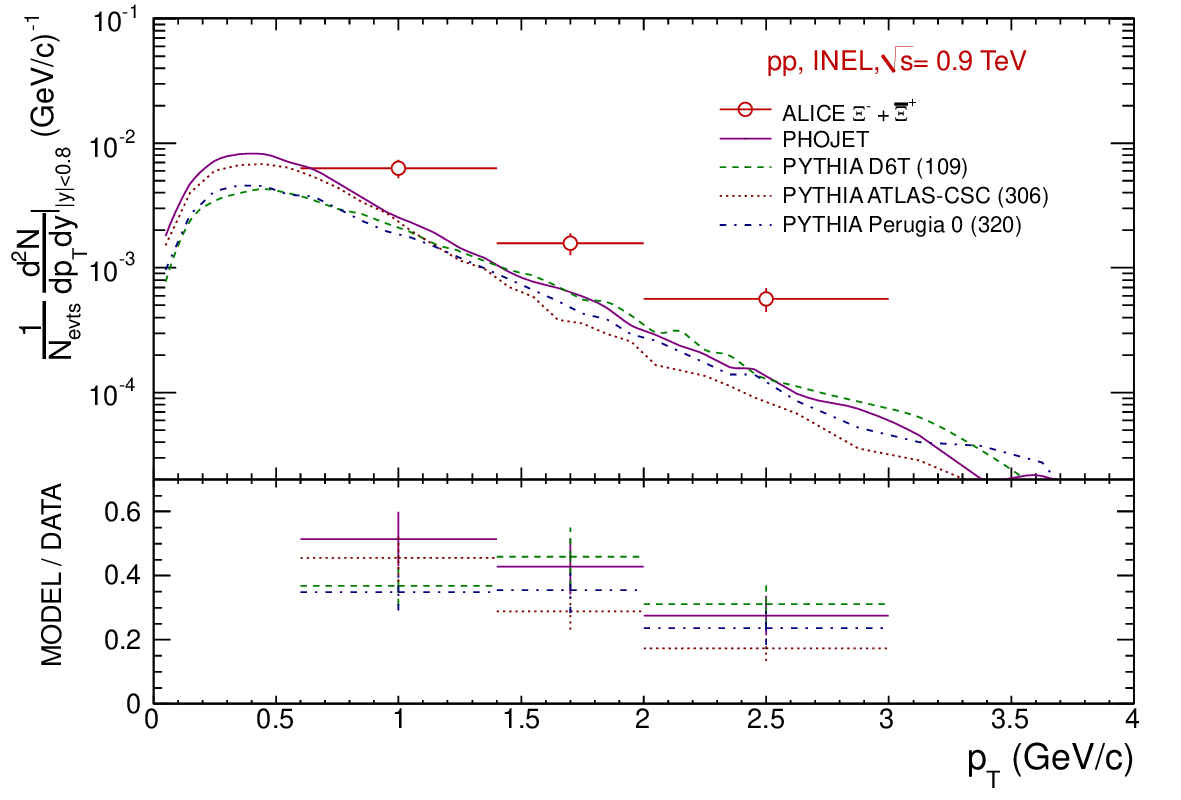}
}
\caption{Comparison of the transverse momentum differential yield for the $\Xis$
particle for INEL pp collisions with PHOJET and PYTHIA tunes 109, 306 and 320.
}
\label{fig:xismc}
\end{figure}

The $\pT$ spectra for $\Kzs$, $\rmLambda$, $\rmAlambda$ and $\phi$ are shown
in Fig.~\ref{fig:corryield} along with the L\'{e}vy fits.
When comparing the different spectra, it is found that the inverse slope parameter
$T$ increases with the mass of the particle.
For example, it changes from $168 \pm 5~\mev$  for $\Kzs$ to $229 \pm 15~\mev$
for $\rmLambda$ when the L\'{e}vy  fit is used.
The $\Xis$ apparently do not follow this trend.
However, this is most likely because the very limited statistics do not allow for a
well-constrained fit.
The shapes of the $\pT$ spectra are also compared to PHOJET and PYTHIA models.
For PYTHIA, several tunes (109~\cite{Albrow:2006rt}, 306~\cite{Moraes:306AIN}
and 320~\cite{Skands:2009zm}) are presented.
For all species, the $\pT$ spectra are found to be slightly harder (i.e. they have a
slower decrease with $\pT$) than the models as presented in Figs.~\ref{fig:k0smc},
\ref{fig:lammc}, \ref{fig:phimc} and \ref{fig:xismc}.
For transverse momenta larger than $\sim 1~\gmom$, the strange particle spectra
are strongly underestimated by all models, by a factor of $\sim 2$ for $\Kzs$ and even
$\sim 3$ for hyperons.
The discrepancy is smaller in the case of the $\phi$.

The integrated yields ($\dNdy$) are obtained using the spectra in the measured range and
integrating the L\'{e}vy function for the extrapolated regions at low and high $\pT$.
The uncertainties for the $\dNdy$ and $\meanpT$ values are computed from the errors 
on the fit parameters, where both the point-to-point statistical and systematic uncertainties of the $\pT$ spectra
are taken into account.
Due to the rapid decrease of the spectra, most of the extrapolation is done in the
low $\pT$ region and amounts to $12~\%$ for $\Kzs$ and $48~\%$ for the $\phi$
(smallest and highest values respectively).
%
%
\begin{table*}
\caption{Rapidity and $\pT$ ranges, $\meanpT$, corrected yields and extrapolated fraction at low $\pT$ using the L\'{e}vy function~(\ref{eqn:funclevy}).
}
\label{tab:6}       
\begin{center}
\resizebox{1.00\textwidth}{!}{
\begin{tabular}{lcccccc}
\hline\noalign{\smallskip}
\multicolumn{2}{c}{Particles}            & $\rap$   & $\pT$ range ($\gmom$)  & $\meanpT$ ($\gmom$)      & $\dNdy$                       & Extrapolation ($\%$) \\
\noalign{\smallskip}\hline\noalign{\smallskip}
\multirow{2}{*}{Mesons}   & $\Kzs$       & $<0.75$  & $[0.2-3.0]$            & $0.65 \pm 0.01 \pm 0.01$ & $0.184 \pm 0.002 \pm 0.006$   & $12 \pm  0.4 \pm 0.5 $   \\
                          & $\phi$       & $<0.60$  & $[0.7-3.0]$            & $1.00 \pm 0.14 \pm 0.20$ & $0.021 \pm 0.004 \pm 0.003$   & $48 \pm 18   \pm 7$   \\
\noalign{\smallskip}\hline\noalign{\smallskip}
\multirow{3}{*}{Baryons}  & $\rmLambda$  & $<0.75$  & $[0.6-3.5]$            & $0.86 \pm 0.01 \pm 0.01$ & $0.048  \pm 0.001 \pm 0.004$  & $36 \pm  2   \pm 4$   \\
                          & $\rmAlambda$ & $<0.75$  & $[0.6-3.5]$            & $0.84 \pm 0.02 \pm 0.02$ & $0.047  \pm 0.002 \pm 0.005$  & $39 \pm  3   \pm 4$   \\
                          & $\Xis$       & $<0.8 $  & $[0.6-3.0]$            & $0.95 \pm 0.14 \pm 0.03$ & $0.0101 \pm 0.0020 \pm 0.0009$ & $35 \pm 8   \pm 4$   \\
\noalign{\smallskip}\hline
\end{tabular}
}
\end{center}
\end{table*}

\begin{figure}
\resizebox{0.5\textwidth}{!}{%
  \includegraphics{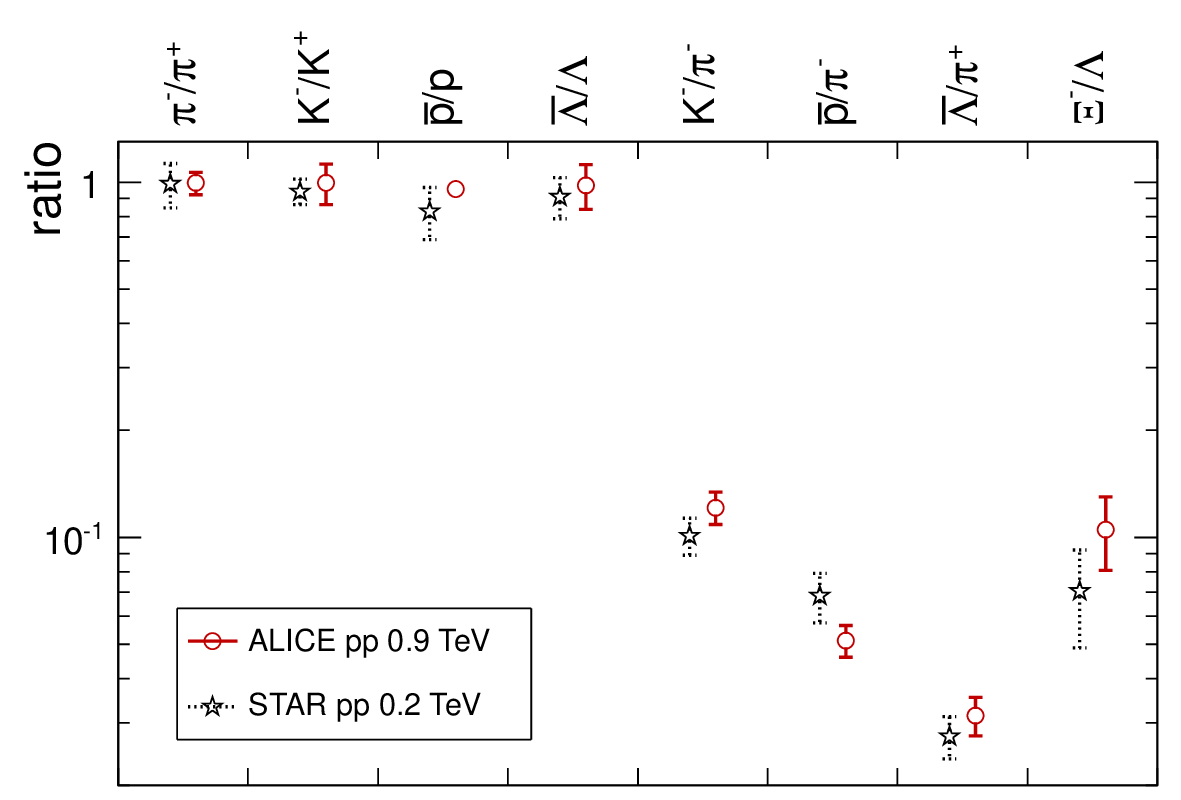}
}
\caption{Ratios of integrated yields including $\pi^{(\pm)}$, K$^{(\pm)}$, p and $\pbar$
performed with the ALICE experiment~\cite{Alice:Pid,Aamodt:2010dx} and compared
with STAR values for pp collisions at $\sqrt{s}$~=~0.2~$\tev$~\cite{Abelev:2006cs}.
All ratios are feed-down corrected.
For the ratio $\rmXi / \rmLambda$ of ALICE, the $\dNdy|_{y=0}$ for $\Xis$ is divided by 2.
Statistical and systematic uncertainties are added in quadrature.}
\label{fig:asrat}
\end{figure}

Therefore, an additional uncertainty is added for the $\dNdy$ to account for the uncertainty
in the shape of the spectra outside the measured range: it corresponds to $25\%$ of the extrapolated particle
yields at low $\pT$.
The measured $\pT$ ranges are specified in Table~\ref{tab:6} for each particle species.
Using the particle integrated yields presented in this paper along with the yields
of charged $\pi$, K, p and $\pbar$~\cite{Alice:Pid} and the measured $\pbar$/p
ratio~\cite{Aamodt:2010dx}, a comparison with STAR feed-down corrected particle
ratios at $\sqrt{s}$~=~0.2~$\tev$~\cite{Abelev:2006cs} is shown in Fig.~\ref{fig:asrat}.
With the centre of mass energy increasing from $\sqrt{s}$~=~0.2~$\tev$ to 0.9~$\tev$
the measured ratios are similar except the $\bar{p}/\pi^{-}$ ratio which decreases
slightly from $0.068 \pm 0.011$ to $0.051 \pm 0.005$.
The strange to non-strange particle ratios seem to increase but stay compatible
within uncertainties:
the $\rm{K}^{-}/\pi^{-}$ from $0.101 \pm 0.012$ to $0.121 \pm 0.013$
and the $\rmAlambda/\pi^{+}$ from $0.027 \pm 0.004$ to $0.032 \pm 0.003$.
\begin{figure}
\resizebox{0.5\textwidth}{!}{%
  \includegraphics{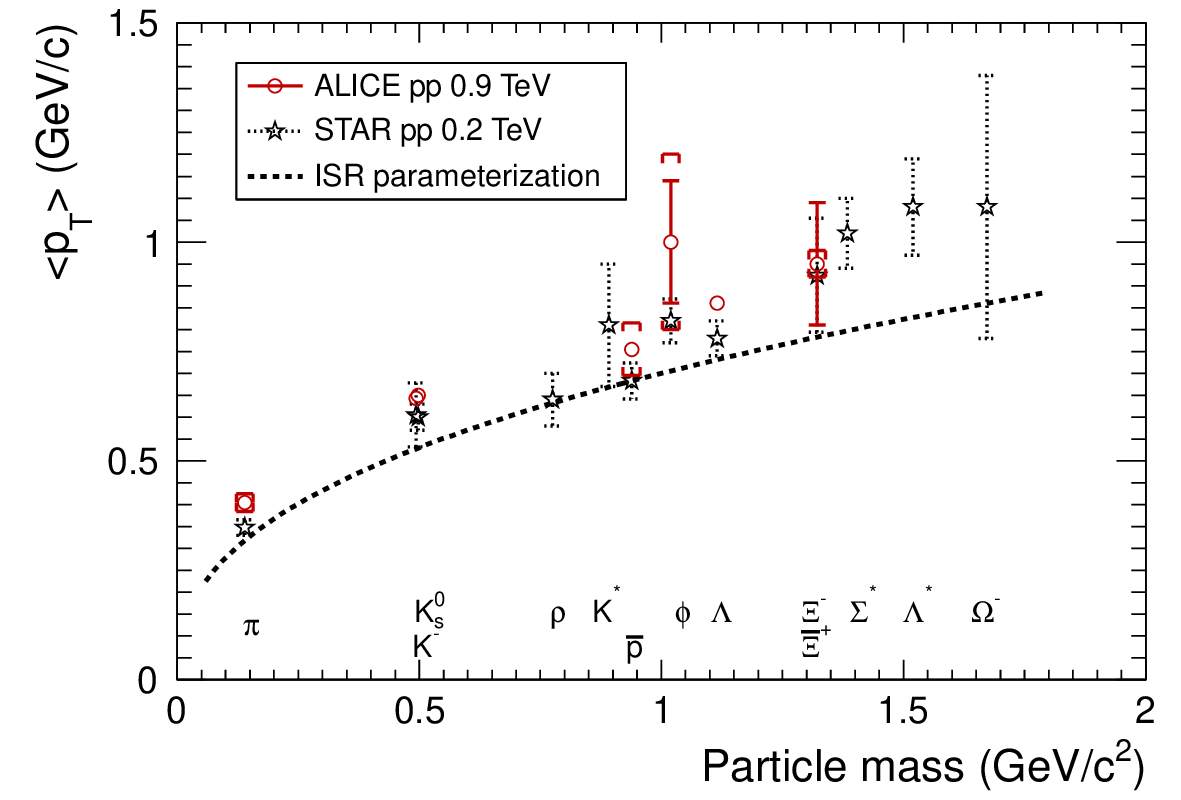}
}
\caption{$\meanpT$ vs. particle mass for the measurements performed with the ALICE experiment
and compared with STAR values for pp collisions at $\sqrt{s}$~=~0.2~$\tev$~\cite{Abelev:2006cs,Adams:2006yu}
and the ISR parameterization~\cite{Bourquin:1976fe}.
Both statistical (vertical error bars) and systematic (brackets) uncertainties are shown for ALICE data. 
}
\label{fig:meanpt}
\end{figure}

The yields and $\meanpT$ obtained with the ALICE experiment are compared for each
particle with existing data at the same energy and also with results at
lower and higher energies.
The various experiments differ in acceptance and event selection (i.e. NSD or INEL) but
the dependence of $\meanpT$ with respect to these variables is found to be negligible.
Consequently the $\meanpT$ values are directly comparable, whereas the comparison
of the yields can require further scaling because of different (pseudo)rapidity coverages. 
Figure~\ref{fig:meanpt} reports ALICE $\meanpT$ measurements along with those of the
STAR experiment~\cite{Abelev:2006cs,Adams:2006yu}.
It is remarkable that the $\meanpT$ remains close to the ISR parameterization~\cite{Bourquin:1976fe}
although the collision energy increased by a factor $36$.
Table~\ref{tab:7} summarizes the $\Kzs$ measurements performed by the
UA5~\cite{Ansorge:1988fq}, CDF~\cite{Abe:1989hy} and ALICE Collaborations for INEL
events, and by the STAR~\cite{Abelev:2006cs} Collaboration for NSD events.
The ALICE $\Kzs$ yield at central rapidity, as well as the $\meanpT$, are in good agreement
with UA5 results at $900~\gev$ albeit with improved precision.
The comparison of $(\rmLambda+\rmAlambda)$ measurements are presented in Table~\ref{tab:8}
for NSD events.
ALICE yields, measured in $\rap<0.75$ for INEL events, are scaled to the UA5~\cite{Alner:1987wb,Ansorge:1989ba}
acceptance ($\rap<2.0$) using PYTHIA simulations.
The $\rmLambda+\rmAlambda$ yield in NSD events is estimated by scaling the measured
yield in inelastic events with the known ratio R of charged particle multiplicities in NSD and
INEL events:
$$ R  = \frac{(\dNdy)_{N_{ch} \rm{NSD}}} {(\dNdy)_{N_{ch} \rm{INEL}}}  = 0.830 \pm 0.024 $$
This scaling factor is also used for the ALICE ($\rmXi+\rmAxi$) yield presented in Table~\ref{tab:9}.
The ALICE yields and $\meanpT$ for both ($\rmLambda+\rmAlambda$) and ($\rmXi+\rmAxi$) are
in good agreement with the UA5 measurements~\cite{Ansorge:1989ba}.
Table~\ref{tab:10} shows the evolution of $\dNdy$ and $\meanpT$ with the collision energy for
the $\phi$ particle in NSD events.
It includes the ALICE measurements, which are the first $\phi$ measurements
at $900~\gev$, and compares them to the results from the STAR experiment~\cite{Adams:2004ux,:2008fd}
at $200~\gev$ and the E735 experiment~\cite{Alexopoulos:1995ru} at $1800~\gev$.
%
%
\begin{table*}
\caption{The $\Kzs$ mean transverse momentum and yields in INEL events from UA5, CDF, and
ALICE and in NSD events in STAR for various $\sqrt{s}$.
STAR results are taken from~\cite{Abelev:2006cs}, CDF ones and yield values with ``$^{\rm *}$" are 
from~\cite{Abe:1989hy}. Other UA5 values concerning $\meanpT$ are from~\cite{Ansorge:1988fq}.}
\label{tab:7}       
\begin{center}
\begin{tabular}{lcccc}
\hline\noalign{\smallskip}
Experiment    &  $\sqrt{s}$ ($\gev$) & acceptance & $\meanpT$ ($\gmom$)   & $\dNdy\vert_{y=0}$ \\
\noalign{\smallskip}\hline\noalign{\smallskip}
 STAR     & $ 200$ &  $\rap<0.5 $ & $0.61 \pm 0.02$            & $0.134 \pm 0.011$ \\
 UA5      & $ 200$ &  $\rap<2.5 $  & $0.53 \pm 0.07$            & $0.14 \pm 0.02^{\rm *}$ \\
 UA5      & $ 546$ &  $\rap<2.5 $  & $0.57 \pm 0.03$            & $0.15 \pm 0.02^{\rm *}$ \\
 CDF      & $ 630$ &  $\rap<1.0 $  & $0.5  \pm 0.1$             & $0.2 \pm 0.1^{\rm *}$ \\
 UA5      & $ 900$ &  $\rap<2.5 $  & $0.62 \pm 0.08$            & $0.18 \pm 0.02^{\rm *}$ \\
 ALICE    & $ 900$ &  $\rap<0.75$  & $0.65 \pm 0.01 \pm 0.01$   & $0.184 \pm 0.002 \pm 0.006$ \\
 CDF      & $1800$ &  $\rap<1.0$  & $0.60 \pm 0.03$            & $0.26 \pm 0.03^{\rm *}$ \\
\noalign{\smallskip}\hline
\end{tabular}
\end{center}
\end{table*}

%
%
\begin{table*}
\caption{The ($\rmLambda+ \rmAlambda$) mean transverse momentum and yields for NSD events and
different $\sqrt{s}$.
STAR results are from~\cite{Abelev:2006cs} and UA5 results are from~\cite{Alner:1987wb,Ansorge:1989ba}.
ALICE and STAR results are feed-down corrected.
The yield measured by ALICE has been scaled to match UA5 acceptance ($\rap<2.0$) using the method
explained in section~\ref{sec_res}.} 
\label{tab:8}       
\begin{center}
\resizebox{1.00\textwidth}{!}{
\begin{tabular}{lcccccc}
\hline\noalign{\smallskip}
Experiment    &  $\sqrt{s}$ ($\gev$) & acceptance & $\meanpT$ ($\gmom$)   & $\dNdy\vert_{y=0}$    &\multicolumn{2}{c}{$\langle n_{\rmLambda+\rmAlambda}\rangle$ per event} \\
              &                      &            &                       &            & measured        & scaled to UA5 $\rap$ \\
\noalign{\smallskip}\hline\noalign{\smallskip}
STAR     & $ 200$ &  $\rap<0.5 $  & $0.77 \pm 0.04$              & $0.074 \pm 0.005$  & ---             & $0.24 \pm 0.02$ \\
UA5      & $ 200$ &  $\rap<2.0 $  & $0.80 ^{+0.20}_{-0.14}$      & ---                 & $0.27 \pm 0.07$ &---          \\
UA5      & $ 546$ &  $\rap<2.0 $  & $0.62 \pm 0.08$              & ---                 & $0.25 \pm 0.05$ &---           \\
UA5      & $ 900$ &  $\rap<2.0 $  & $0.74 \pm 0.09$              & ---                 & $0.38 \pm 0.08$ &---           \\
ALICE    & $ 900$ &  $\rap<0.75$  & $0.85 \pm 0.01 \pm 0.01$     & $0.095 \pm 0.002 \pm 0.003$  & ---     &  $0.46 \pm 0.01 \pm 0.02$\\
\noalign{\smallskip}\hline
\end{tabular}
}
\end{center}
\end{table*}

%
%
\begin{table*}
\caption{The ($\rmXi+\rmAxi$) mean transverse momentum and yields for NSD events and different $\sqrt{s}$.
STAR results are from~\cite{Abelev:2006cs} and UA5 results are from~\cite{Ansorge:1989ba}.
UA5 measures ($\rmXi+\rmAxi$) for $\pT > 1~\gmom$.
The ALICE yield has been scaled to match the UA5 acceptance ($\rap<3.0$) using the method explained in
section~\ref{sec_res}.}
\label{tab:9}       
\begin{center}
\resizebox{1.00\textwidth}{!}{
\begin{tabular}{lcccccc}
\hline\noalign{\smallskip}
Experiment    &  $\sqrt{s}$ ($\gev$) & acceptance & $\meanpT$ ($\gmom$)   &  $\dNdy\vert_{y=0}$    &\multicolumn{2}{c}{$\langle n_{\rmXi+\rmAxi}\rangle$ per event} \\
              &                      &            &                       &            & measured        & scaled to UA5 $\rap$ \\
\noalign{\smallskip}\hline\noalign{\smallskip}
STAR     & $ 200$ &  $\rap<0.5 $   & $0.90 \pm 0.01$         & $0.006 \pm 0.001$           &---                          & $0.022 \pm 0.006$ \\
UA5      & $ 200$ &  $\rap<3.0 $   & $0.80^{+0.20}_{-0.14}$    & ---                           & $0.03^{+0.04}_{-0.02}$       & --- \\
UA5      & $ 546$ &  $\rap<3.0 $   & $1.10 \pm 0.02$            & ---                           & $0.08^{+0.03}_{-0.02}$       & --- \\
UA5      & $ 900$ &  $\rap<3.0 $   & $0.7^{+0.2}_{-0.1}$       & ---                           & $0.05^{+0.04}_{-0.02}$       & --- \\
ALICE    & $ 900$ &  $\rap<0.8$    & $0.95 \pm 0.14 \pm 0.03$  & $0.0101 \pm 0.0020 \pm 0.0009$ &  &  $0.078 \pm  0.015 \pm 0.007$\\

\noalign{\smallskip}\hline
\end{tabular}
}
\end{center}
\end{table*}

%
%
\begin{table*}
\caption{The $\phi$ mean transverse momentum and yields for NSD events and different $\sqrt{s}$.
STAR results are from~\cite{Adams:2004ux,:2008fd} and E735 results are from~\cite{Alexopoulos:1995ru}.
The E735 Collaboration provided two values of $\meanpT$ depending on the functional form used to fit
the data points and the uncertainties associated with each value are only statistical.
ALICE yields measured for INEL events have been scaled to NSD as explained in section~\ref{sec_res}.}
\label{tab:10}       
\begin{center}
\begin{tabular}{lcccc}
\hline\noalign{\smallskip}
Experiment    &  $\sqrt{s}$ ($\gev$) & acceptance & $\meanpT$ ($\gmom$)  & $\dNdy\vert_{y=0}$ \\
\noalign{\smallskip}\hline\noalign{\smallskip}
STAR     & $ 200$ &  $\rap<0.5 $         & $0.82 \pm 0.03 \pm 0.04$   & $0.018 \pm 0.001 \pm 0.003$ \\
ALICE    & $ 900$ &  $\rap<0.6$          & $1.00 \pm 0.14 \pm 0.20$   & $0.021 \pm 0.004 \pm 0.003$  \\
\multirow{2}{*}{E735} & \multirow{2}{*}{$1800$} &  \multirow{2}{*}{$-0.4 <y< 1.0 $}  & $1.06 \pm 0.18$  & \multirow{2}{*}{$0.0186 \pm 0.0041$}  \\
         &        &                      & $0.94 \pm 0.26$            &                      \\
\noalign{\smallskip}\hline
\end{tabular}
\end{center}
\end{table*}

\begin{figure}
\resizebox{0.5\textwidth}{!}{%
  \includegraphics{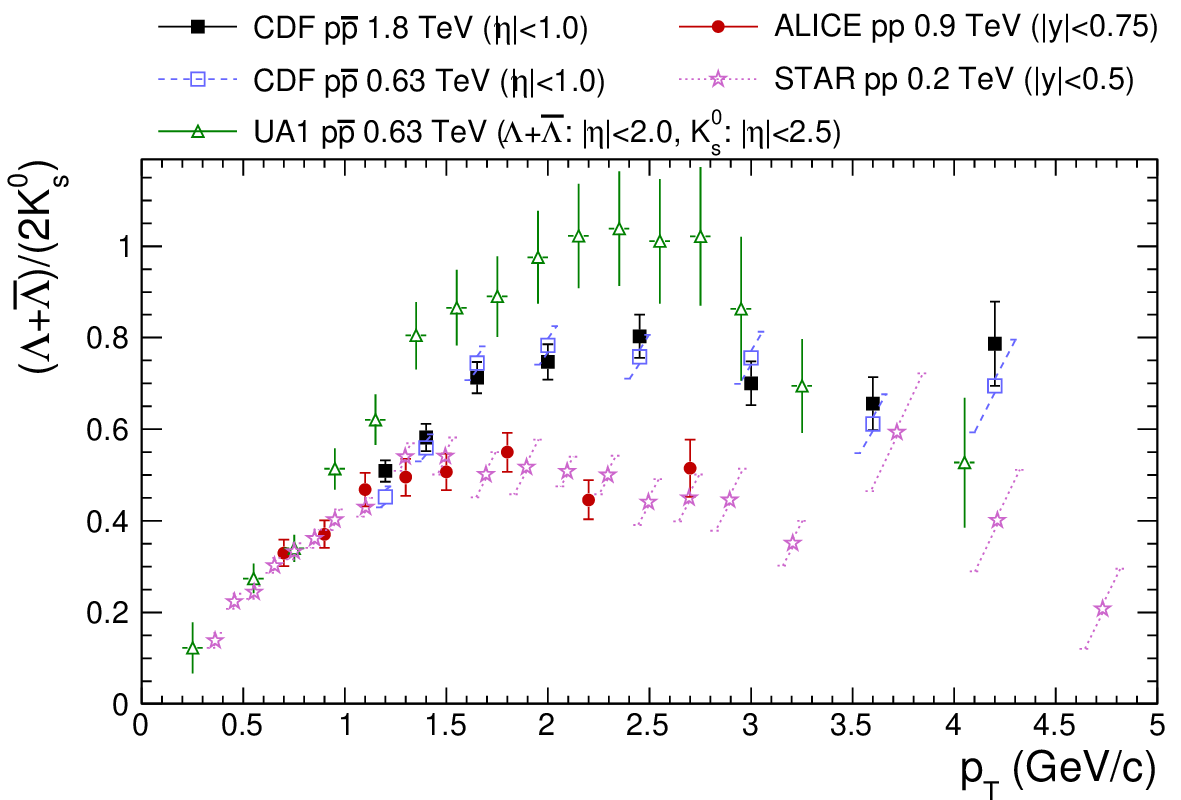}
}
\caption{$(\rmLambda+\rmAlambda)/2\Kzs$ as a function of $\pT$ for different collision energies
in pp and $\ppbar$ minimum bias events.
The STAR ratio is taken from~\cite{Abelev:2006cs} whereas the CDF and UA1 ratios are computed
with the $(\rmLambda+\rmAlambda)$ and $\Kzs$ spectra published  in~\cite{Acosta:2005pk}
and~\cite{Bocquet:1995jq} respectively.
The ALICE and STAR ratios are feed-down corrected.
Because the $\Kzs$ and $(\rmLambda+\rmAlambda)$ spectra from UA1 have incompatible binning,
the $\Kzs$ differential yield has been calculated for each $(\rmLambda+\rmAlambda)$ 
$\pT$ data point using the fit function published by UA1. Such a choice is motivated by the fact 
that the $\chi^{2}$ value for the $\Kzs$ spectrum fit is better than that for the
$(\rmLambda+\rmAlambda)$ spectrum.
}
\label{fig:lamk0s}
\end{figure}

The baryon to meson ratio as a function of $\pT$ obtained with the $(\rmLambda+\rmAlambda)$
and $\Kzs$ spectra measured by ALICE is presented in Fig.~\ref{fig:lamk0s}.
It includes the $(\rmLambda+\rmAlambda)/2\Kzs$ ratio in pp collisions at $200~\gev$ measured by
STAR~\cite{Abelev:2006cs}, and the ratios in $\ppbar$ collisions at $630~\gev$ and $1800~\gev$
computed with the $(\rmLambda+\rmAlambda)$ and $\Kzs$ spectra published by CDF~\cite{Acosta:2005pk}
and UA1~\cite{Bocquet:1995jq}.
UA1 and CDF Collaborations provide inclusive spectra.
The associated ratios are therefore not feed-down corrected, unlike the ALICE and STAR ones. 
The acceptance windows of these experiments differ significantly:
ALICE measures $\rmLambda$, $\rmAlambda$ and $\Kzs$ in $\rap < 0.75$, STAR in $\rap < 0.5$,
CDF in $\pseudorap < 1.0$, whereas UA1 reconstructs $(\rmLambda + \rmAlambda)$ in
$\pseudorap < 2.0$ and $\Kzs$ in $\pseudorap < 2.5$. 
The ALICE ratio agrees very well with the STAR results in the measured $\pT$ range,
which would suggest little or no energy dependence of $(\rmLambda+\rmAlambda)/2\Kzs$.
A similar conclusion can be drawn when comparing only the ratios measured by CDF at $630~\gev$
and $1800~\gev$, although the ratio found by CDF for $\pT > 1.5~\gmom$ is higher than the
one observed with ALICE and STAR.
The ratio computed from UA1 spectra however shows a clear disagreement with the other measurements
in an intermediate $\pT$ range between $\pT \approx 1.5~\gmom$ and $\pT \approx 3.0~\gmom$.
PYTHIA simulations show that this discrepancy can not be attributed to the differences in
the acceptance or in the colliding system (i.e. $\ppbar$ instead of pp).
\section{Conclusions}
Measurements of mesons containing strange quarks ($\Kzs$ and $\phi$)
and hyperons ($\rmLambda$, $\rmAlambda$ and $\Xis$) have been performed
for inelastic pp collisions at $\sqrt{s}= 0.9~\tev$ with the ALICE experiment at the LHC.
The L\'{e}vy function gives a good description of the transverse momentum
spectra which have been compared  with pQCD-based models.
The $\Kzs$ transverse momentum spectrum is overestimated by PYTHIA tune 
ATLAS-CSC and PHOJET below $0.75~\gmom$ but is higher by a factor of 
$\sim 2$ in the $\pT$ range $[1-3]~\gmom$.
Within uncertainties, the $\phi$ meson spectrum is reasonably described by
these models and the best agreement is obtained by PYTHIA tune D6T.
We find that strange baryons are significantly under-predicted in both
PYTHIA and PHOJET by a factor of $\sim 3$.
The feed-down corrected ratio of baryon to meson as a function of $\pT$,
illustrated by the $\rmLambda/\Kzs$, is consistent with the STAR measurements
at $\sqrt{s}= 0.2~\tev$ but lower than UA1 and CDF results at $\sqrt{s}= 0.63~\tev$
and $\sqrt{s}= 1.8~\tev$.
The integrated yields and average transverse momenta have been compared
with earlier data collected in pp and $\ppbar$ interactions at various energies.
These results provide a useful baseline for comparisons with recent tunes
of the PYTHIA model and a reference for future measurements in heavy-ion
collisions at the LHC.
These studies demonstrate the precision with which ALICE can measure
resonances and topologically reconstructed weakly decaying particles.
Measurements of these particles will be a substantial part of the ALICE
programme in both pp and \PbPb collisions.
The measurement of the $\phi$ resonance provides an unprecedented
reference at this energy.
%
%
%
\newenvironment{acknowledgement}{\relax}{\relax}
\begin{acknowledgement}
\section*{Acknowledgements}
The ALICE collaboration would like to thank all its engineers and technicians for their invaluable contributions to the construction of the experiment and the CERN accelerator teams for the outstanding performance of the LHC complex.
The ALICE collaboration acknowledges the following funding agencies for their support in building and
running the ALICE detector:
Calouste Gulbenkian Foundation from Lisbon and Swiss Fonds Kidagan, Armenia;
Conselho Nacional de Desenvolvimento Cient\'{\i}fico e Tecnol\'{o}gico (CNPq), Financiadora de Estudos e Projetos (FINEP),
Funda\c{c}\~{a}o de Amparo \`{a} Pesquisa do Estado de S\~{a}o Paulo (FAPESP);
National Natural Science Foundation of China (NSFC), the Chinese Ministry of Education (CMOE)
and the Ministry of Science and Technology of China (MSTC);
Ministry of Education and Youth of the Czech Republic;
Danish Natural Science Research Council, the Carlsberg Foundation and the Danish National Research Foundation;
The European Research Council under the European Community's Seventh Framework Programme;
Helsinki Institute of Physics and the Academy of Finland;
French CNRS-IN2P3, the `Region Pays de Loire', `Region Alsace', `Region Auvergne' and CEA, France;
German BMBF and the Helmholtz Association;
Greek Ministry of Research and Technology;
Hungarian OTKA and National Office for Research and Technology (NKTH);
Department of Atomic Energy and Department of Science and Technology of the Government of India;
Istituto Nazionale di Fisica Nucleare (INFN) of Italy;
MEXT Grant-in-Aid for Specially Promoted Research, Ja\-pan;
Joint Institute for Nuclear Research, Dubna;
 %
National Research Foundation of Korea (NRF);
CONACYT, DGAPA, M\'{e}xico, ALFA-EC and the HELEN Program (High-Energy physics Latin-American--European Network);
Stichting voor Fundamenteel Onderzoek der Materie (FOM) and the Nederlandse Organisatie voor Wetenschappelijk Onderzoek (NWO), Netherlands;
Research Council of Norway (NFR);
Polish Ministry of Science and Higher Education;
National Authority for Scientific Research - NASR (Autoritatea Na\c{t}ional\u{a} pentru Cercetare \c{S}tiin\c{t}ific\u{a} - ANCS);
Federal Agency of Science of the Ministry of Education and Science of Russian Federation, International Science and
Technology Center, Russian Academy of Sciences, Russian Federal Agency of Atomic Energy, Russian Federal Agency for Science and Innovations and CERN-INTAS;
Ministry of Education of Slovakia;
CIEMAT, EELA, Ministerio de Educaci\'{o}n y Ciencia of Spain, Xunta de Galicia (Conseller\'{\i}a de Educaci\'{o}n),
CEA\-DEN, Cubaenerg\'{\i}a, Cuba, and IAEA (International Atomic Energy Agency);
The Ministry of Science and Technology and the National Research Foundation (NRF), South Africa;
Swedish Reseach Council (VR) and Knut $\&$ Alice Wallenberg Foundation (KAW);
Ukraine Ministry of Education and Science;
United Kingdom Science and Technology Facilities Council (STFC);
The United States Department of Energy, the United States National
Science Foundation, the State of Texas, and the State of Ohio.
 
\end{acknowledgement}
\newpage
%
\appendix
\section{The ALICE Collaboration}
\label{app:collab}
%
\begingroup
\small
\begin{flushleft}
K.~Aamodt\Irefn{0}\And
A.~Abrahantes~Quintana\Irefn{1}\And
D.~Adamov\'{a}\Irefn{2}\And
A.M.~Adare\Irefn{3}\And
M.M.~Aggarwal\Irefn{4}\And
G.~Aglieri~Rinella\Irefn{5}\And
A.G.~Agocs\Irefn{6}\And
S.~Aguilar~Salazar\Irefn{7}\And
Z.~Ahammed\Irefn{8}\And
N.~Ahmad\Irefn{9}
% ...
\renewcommand\labelenumi{\textsuperscript{\theenumi}~}
\section*{Affiliation notes}
\renewcommand\theenumi{\roman{enumi}}
\begin{Authlist}
\item \Adef{0}Also at Laboratoire de Physique Corpusculaire (LPC), Clermont Universit\'{e}, Universit\'{e} Blaise Pascal, CNRS--IN2P3, Clermont-Ferrand, France
\item \Adef{1}Now at Centro Fermi -- Centro Studi e Ricerche e Museo Storico della Fisica ``Enrico Fermi'', Rome, Italy
\item \Adef{2}Now at Physikalisches Institut, Ruprecht-Karls-Universit\"{a}t Heidelberg, Heidelberg, Germany
\item \Adef{3}Now at Frankfurt Institute for Advanced Studies, Johann Wolfgang Goethe-Universit\"{a}t Frankfurt, Frankfurt, Germany
\item \Adef{4}Now at Sezione INFN, Turin, Italy
\item \Adef{5}Now at University of Houston, Houston, Texas, United States
\item \Adef{6}Also at  Dipartimento di Fisica dell'Universit\'{a}, Udine, Italy 
\item \Adef{7}Now at SUBATECH, Ecole des Mines de Nantes, Universit\'{e} de Nantes, CNRS-IN2P3, Nantes, France
\item \Adef{8}Now at Centro de Investigaci\'{o}n y de Estudios Avanzados (CINVESTAV), Mexico City and M\'{e}rida, Mexico
\item \Adef{9}Now at Benem\'{e}rita Universidad Aut\'{o}noma de Puebla, Puebla, Mexico
\item \Adef{10}Now at Laboratoire de Physique Subatomique et de Cosmologie (LPSC), Universit\'{e} Joseph Fourier, CNRS-IN2P3, Institut Polytechnique de Grenoble, Grenoble, France
\item \Adef{11}Now at Institut Pluridisciplinaire Hubert Curien (IPHC), Universit\'{e} de Strasbourg, CNRS-IN2P3, Strasbourg, France
\item \Adef{12}Now at Sezione INFN, Padova, Italy
\item \Adef{13} Deceased 
\item \Adef{14}Also at Division of Experimental High Energy Physics, University of Lund, Lund, Sweden
\item \Adef{15}Also at  University of Technology and Austrian Academy of Sciences, Vienna, Austria 
\item \Adef{16}Also at European Organization for Nuclear Research (CERN), Geneva, Switzerland
\item \Adef{17}Now at Oak Ridge National Laboratory, Oak Ridge, Tennessee, United States
\item \Adef{18}Now at European Organization for Nuclear Research (CERN), Geneva, Switzerland
\item \Adef{19}Also at Wayne State University, Detroit, Michigan, United States
\item \Adef{20}Also at Frankfurt Institute for Advanced Studies, Johann Wolfgang Goethe-Universit\"{a}t Frankfurt, Frankfurt, Germany
\item \Adef{21}Now at Research Division and ExtreMe Matter Institute EMMI, GSI Helmholtzzentrum f\"ur Schwerionenforschung, Darmstadt, Germany
\item \Adef{22}Also at Fachhochschule K\"{o}ln, K\"{o}ln, Germany
\item \Adef{23}Also at Institute of Experimental Physics, Slovak Academy of Sciences, Ko\v{s}ice, Slovakia
\item \Adef{24}Now at Instituto de Ciencias Nucleares, Universidad Nacional Aut\'{o}noma de M\'{e}xico, Mexico City, Mexico
\item \Adef{25}Also at Laboratoire de Physique Subatomique et de Cosmologie (LPSC), Universit\'{e} Joseph Fourier, CNRS-IN2P3, Institut Polytechnique de Grenoble, Grenoble, France
\item \Adef{26}Also at  "Vin\v{c}a" Institute of Nuclear Sciences, Belgrade, Serbia 
\item \Adef{27}Also at University of Houston, Houston, Texas, United States
\item \Adef{28}Also at Department of Physics, University of Oslo, Oslo, Norway
\item \Adef{29}Also at Variable Energy Cyclotron Centre, Kolkata, India
\item \Adef{30}Now at Department of Physics, University of Oslo, Oslo, Norway
\item \Adef{31}Also at Dipartimento Interateneo di Fisica `M.~Merlin' and Sezione INFN, Bari, Italy
\item \Adef{32}Now at Nikhef, National Institute for Subatomic Physics and Institute for Subatomic Physics of Utrecht University, Utrecht, Netherlands
\item \Adef{33}Also at Hua-Zhong Normal University, Wuhan, China
\item \Adef{34}Also at Centro Fermi -- Centro Studi e Ricerche e Museo Storico della Fisica ``Enrico Fermi'', Rome, Italy
\end{Authlist}
\section*{Collaboration Institutes}
\renewcommand\theenumi{\arabic{enumi}~}
\begin{Authlist}
\item \Idef{0}Department of Physics and Technology, University of Bergen, Bergen, Norway
\item \Idef{1}Centro de Aplicaciones Tecnol\'{o}gicas y Desarrollo Nuclear (CEADEN), Havana, Cuba
\item \Idef{2}Nuclear Physics Institute, Academy of Sciences of the Czech Republic, \v{R}e\v{z} u Prahy, Czech Republic
\item \Idef{3}Yale University, New Haven, Connecticut, United States
\item \Idef{4}Physics Department, Panjab University, Chandigarh, India
\item \Idef{5}European Organization for Nuclear Research (CERN), Geneva, Switzerland
\item \Idef{6}KFKI Research Institute for Particle and Nuclear Physics, Hungarian Academy of Sciences, Budapest, Hungary
\item \Idef{7}Instituto de F\'{\i}sica, Universidad Nacional Aut\'{o}noma de M\'{e}xico, Mexico City, Mexico
\item \Idef{8}Variable Energy Cyclotron Centre, Kolkata, India
\item \Idef{9}Department of Physics Aligarh Muslim University, Aligarh, India
\item \Idef{10}Gangneung-Wonju National University, Gangneung, South Korea
\item \Idef{11}Institute for Theoretical and Experimental Physics, Moscow, Russia
\item \Idef{12}Russian Research Centre Kurchatov Institute, Moscow, Russia
\item \Idef{13}Sezione INFN, Turin, Italy
\item \Idef{14}Dipartimento di Fisica dell'Universit\`{a} and Sezione INFN, Bologna, Italy
\item \Idef{15}Bogolyubov Institute for Theoretical Physics, Kiev, Ukraine
\item \Idef{16}Frankfurt Institute for Advanced Studies, Johann Wolfgang Goethe-Universit\"{a}t Frankfurt, Frankfurt, Germany
\item \Idef{17}Dipartimento Interateneo di Fisica `M.~Merlin' and Sezione INFN, Bari, Italy
\item \Idef{18}Research Division and ExtreMe Matter Institute EMMI, GSI Helmholtzzentrum f\"ur Schwerionenforschung, Darmstadt, Germany
\item \Idef{19}V.~Fock Institute for Physics, St. Petersburg State University, St. Petersburg, Russia
\item \Idef{20}National Institute for Physics and Nuclear Engineering, Bucharest, Romania
\item \Idef{21}Kirchhoff-Institut f\"{u}r Physik, Ruprecht-Karls-Universit\"{a}t Heidelberg, Heidelberg, Germany
\item \Idef{22}Department of Physics, Ohio State University, Columbus, Ohio, United States
\item \Idef{23}Rudjer Bo\v{s}kovi\'{c} Institute, Zagreb, Croatia
\item \Idef{24}Dipartimento di Fisica dell'Universit\`{a} and Sezione INFN, Padova, Italy
\item \Idef{25}Sezione INFN, Bologna, Italy
\item \Idef{26}SUBATECH, Ecole des Mines de Nantes, Universit\'{e} de Nantes, CNRS-IN2P3, Nantes, France
\item \Idef{27}Institut f\"{u}r Kernphysik, Johann Wolfgang Goethe-Universit\"{a}t Frankfurt, Frankfurt, Germany
\item \Idef{28}Laboratoire de Physique Subatomique et de Cosmologie (LPSC), Universit\'{e} Joseph Fourier, CNRS-IN2P3, Institut Polytechnique de Grenoble, Grenoble, France
\item \Idef{29}Departamento de F\'{\i}sica de Part\'{\i}culas and IGFAE, Universidad de Santiago de Compostela, Santiago de Compostela, Spain
\item \Idef{30}Oak Ridge National Laboratory, Oak Ridge, Tennessee, United States
\item \Idef{31}Helsinki Institute of Physics (HIP) and University of Jyv\"{a}skyl\"{a}, Jyv\"{a}skyl\"{a}, Finland
\item \Idef{32}Sezione INFN, Catania, Italy
\item \Idef{33}Dipartimento di Fisica Sperimentale dell'Universit\`{a} and Sezione INFN, Turin, Italy
\item \Idef{34}Centro Fermi -- Centro Studi e Ricerche e Museo Storico della Fisica ``Enrico Fermi'', Rome, Italy
\item \Idef{35}Commissariat \`{a} l'Energie Atomique, IRFU, Saclay, France
\item \Idef{36}Laboratoire de Physique Corpusculaire (LPC), Clermont Universit\'{e}, Universit\'{e} Blaise Pascal, CNRS--IN2P3, Clermont-Ferrand, France
\item \Idef{37}Institute of Experimental Physics, Slovak Academy of Sciences, Ko\v{s}ice, Slovakia
\item \Idef{38}Dipartimento di Fisica e Astronomia dell'Universit\`{a} and Sezione INFN, Catania, Italy
\item \Idef{39}School of Physics and Astronomy, University of Birmingham, Birmingham, United Kingdom
\item \Idef{40}The Henryk Niewodniczanski Institute of Nuclear Physics, Polish Academy of Sciences, Cracow, Poland
\item \Idef{41}Institut f\"{u}r Kernphysik, Westf\"{a}lische Wilhelms-Universit\"{a}t M\"{u}nster, M\"{u}nster, Germany
\item \Idef{42}Joint Institute for Nuclear Research (JINR), Dubna, Russia
\item \Idef{43}Niels Bohr Institute, University of Copenhagen, Copenhagen, Denmark
\item \Idef{44}Institut Pluridisciplinaire Hubert Curien (IPHC), Universit\'{e} de Strasbourg, CNRS-IN2P3, Strasbourg, France
\item \Idef{45}Wayne State University, Detroit, Michigan, United States
\item \Idef{46}Petersburg Nuclear Physics Institute, Gatchina, Russia
\item \Idef{47}Physics Department, University of Jammu, Jammu, India
\item \Idef{48}Laboratori Nazionali di Frascati, INFN, Frascati, Italy
\item \Idef{49}Faculty of Nuclear Sciences and Physical Engineering, Czech Technical University in Prague, Prague, Czech Republic
\item \Idef{50}Nikhef, National Institute for Subatomic Physics, Amsterdam, Netherlands
\item \Idef{51}Centro de Investigaciones Energ\'{e}ticas Medioambientales y Tecnol\'{o}gicas (CIEMAT), Madrid, Spain
\item \Idef{52}University of Houston, Houston, Texas, United States
\item \Idef{53}Moscow Engineering Physics Institute, Moscow, Russia
\item \Idef{54}Institute for High Energy Physics, Protvino, Russia
\item \Idef{55}Faculty of Science, P.J.~\v{S}af\'{a}rik University, Ko\v{s}ice, Slovakia
\item \Idef{56}Saha Institute of Nuclear Physics, Kolkata, India
\item \Idef{57}Institut de Physique Nucl\'{e}aire d'Orsay (IPNO), Universit\'{e} Paris-Sud, CNRS-IN2P3, Orsay, France
\item \Idef{58}Department of Physics, University of Oslo, Oslo, Norway
\item \Idef{59}Dipartimento di Fisica dell'Universit\`{a} and Sezione INFN, Trieste, Italy
\item \Idef{60}Faculty of Mathematics, Physics and Informatics, Comenius University, Bratislava, Slovakia
\item \Idef{61}Russian Federal Nuclear Center (VNIIEF), Sarov, Russia
\item \Idef{62}Physikalisches Institut, Ruprecht-Karls-Universit\"{a}t Heidelberg, Heidelberg, Germany
\item \Idef{63}Physics Department, University of Cape Town, iThemba Laboratories, Cape Town, South Africa
\item \Idef{64}Hua-Zhong Normal University, Wuhan, China
\item \Idef{65}Secci\'{o}n F\'{\i}sica, Departamento de Ciencias, Pontificia Universidad Cat\'{o}lica del Per\'{u}, Lima, Peru
\item \Idef{66}Physics Department, Creighton University, Omaha, Nebraska, United States
\item \Idef{67}Universit\'{e} de Lyon, Universit\'{e} Lyon 1, CNRS/IN2P3, IPN-Lyon, Villeurbanne, France
\item \Idef{68}Universidade Estadual de Campinas (UNICAMP), Campinas, Brazil
\item \Idef{69}Nikhef, National Institute for Subatomic Physics and Institute for Subatomic Physics of Utrecht University, Utrecht, Netherlands
\item \Idef{70}Division of Experimental High Energy Physics, University of Lund, Lund, Sweden
\item \Idef{71}University of Tsukuba, Tsukuba, Japan
\item \Idef{72}Sezione INFN, Cagliari, Italy
\item \Idef{73}Centro de Investigaci\'{o}n y de Estudios Avanzados (CINVESTAV), Mexico City and M\'{e}rida, Mexico
\item \Idef{74}Benem\'{e}rita Universidad Aut\'{o}noma de Puebla, Puebla, Mexico
\item \Idef{75}Dipartimento di Scienze e Tecnologie Avanzate dell'Universit\`{a} del Piemonte Orientale and Gruppo Collegato INFN, Alessandria, Italy
\item \Idef{76}Instituto de Ciencias Nucleares, Universidad Nacional Aut\'{o}noma de M\'{e}xico, Mexico City, Mexico
\item \Idef{77}Laboratori Nazionali di Legnaro, INFN, Legnaro, Italy
\item \Idef{78}Institute of Space Sciences (ISS), Bucharest, Romania
\item \Idef{79}Institute of Physics, Bhubaneswar, India
\item \Idef{80}Universidade de S\~{a}o Paulo (USP), S\~{a}o Paulo, Brazil
\item \Idef{81}Dipartimento di Fisica `E.R.~Caianiello' dell'Universit\`{a} and Gruppo Collegato INFN, Salerno, Italy
\item \Idef{82}Sezione INFN, Bari, Italy
\item \Idef{83}Dipartimento di Fisica dell'Universit\`{a} and Sezione INFN, Cagliari, Italy
\item \Idef{84}Soltan Institute for Nuclear Studies, Warsaw, Poland
\item \Idef{85}Sezione INFN, Rome, Italy
\item \Idef{86}Faculty of Engineering, Bergen University College, Bergen, Norway
\item \Idef{87}Sezione INFN, Padova, Italy
\item \Idef{88}Institute for Nuclear Research, Academy of Sciences, Moscow, Russia
\item \Idef{89}Sezione INFN, Trieste, Italy
\item \Idef{90}Physics Department, University of Athens, Athens, Greece
\item \Idef{91}Warsaw University of Technology, Warsaw, Poland
\item \Idef{92}Universidad Aut\'{o}noma de Sinaloa, Culiac\'{a}n, Mexico
\item \Idef{93}Technical University of Split FESB, Split, Croatia
\item \Idef{94}Yerevan Physics Institute, Yerevan, Armenia
\item \Idef{95}University of Tokyo, Tokyo, Japan
\item \Idef{96}Department of Physics, Sejong University, Seoul, South Korea
\item \Idef{97}Lawrence Berkeley National Laboratory, Berkeley, California, United States
\item \Idef{98}Indian Institute of Technology, Mumbai, India
\item \Idef{99}Institut f\"{u}r Kernphysik, Technische Universit\"{a}t Darmstadt, Darmstadt, Germany
\item \Idef{100}Yonsei University, Seoul, South Korea
\item \Idef{101}Zentrum f\"{u}r Technologietransfer und Telekommunikation (ZTT), Fachhochschule Worms, Worms, Germany
\item \Idef{102}California Polytechnic State University, San Luis Obispo, California, United States
\item \Idef{103}China Institute of Atomic Energy, Beijing, China
\item \Idef{104}Institute of Physics, Academy of Sciences of the Czech Republic, Prague, Czech Republic
\item \Idef{105}University of Tennessee, Knoxville, Tennessee, United States
\item \Idef{106}Dipartimento di Fisica dell'Universit\`{a} `La Sapienza' and Sezione INFN, Rome, Italy
\item \Idef{107}Hiroshima University, Hiroshima, Japan
\item \Idef{108}Lawrence Livermore National Laboratory, Livermore, California, United States
\item \Idef{109}Budker Institute for Nuclear Physics, Novosibirsk, Russia
\item \Idef{110}Physics Department, University of Rajasthan, Jaipur, India
\item \Idef{111}Purdue University, West Lafayette, Indiana, United States
\item \Idef{112}Centre de Calcul de l'IN2P3, Villeurbanne, France 
\item \Idef{113}Pusan National University, Pusan, South Korea
\end{Authlist}
\endgroup
  
%
%
%

%
%
%
\end{document}